\documentclass[12pt]{article}
\pdfoutput=1

\usepackage{cite}
\usepackage{setspace}
\usepackage{xspace}
\usepackage{color}
\usepackage{amsmath, amsfonts, amssymb, amsthm, upgreek}
\usepackage{caption}
\usepackage{slashed}
\usepackage{subfig}

\usepackage[english]{babel}
\usepackage[applemac]{inputenc}

\usepackage{hyperref}

\usepackage{ifpdf}
\ifpdf
\usepackage[pdftex]{graphicx}
\usepackage{epstopdf}
\else
\usepackage[dvips]{graphicx}
\fi

\parskip=\baselineskip

\newcommand{\beq}{\begin{equation}}
\newcommand{\eeq}{\end{equation}}
\newcommand{\bea}{\begin{eqnarray}}
\newcommand{\eea}{\end{eqnarray}}
\newcommand{\bag}{\begin{align}}
\newcommand{\eag}{\end{align}}

\newcommand{\Eq}[1]{Eq.\!~(\ref{#1})}

\newcommand{\Eqs}[1]{Eqs.\!~(\ref{#1})}

\newcommand{\GeV}{\,\mathrm{GeV}}
\newcommand{\TeV}{\,\mathrm{TeV}}

\newcommand{\nn}{\nonumber}

\newcommand{\ie}{$\textnormal{i.e.}$ }

\newcommand{\Tr}{\mathrm{Tr}}
\newcommand{\Lag}{\mathcal{L}}
\newcommand{\order}[1]{O(#1)}
\newcommand{\Op}{{\mathcal O}}

\newcommand{\Dslash}{\hspace{3pt}\raisebox{1.5pt}{$\slash$} \hspace{-8pt} D}

\newcommand{\BR}{\mathrm{BR}}

\newcommand{\vev}[1]{\langle {#1} \rangle}

\newcommand{\SU}{\mathrm{SU}}
\newcommand{\SO}{\mathrm{SO}}
\newcommand{\UU}{\mathrm{U}}
\newcommand{\Sp}{\mathrm{Sp}}

\newcommand{\cp}{\mathrm{CP}}
\newcommand{\ahat}{i}
\newcommand{\diag}{\mathrm{diag}}
\newcommand{\fpi}{f}
\newcommand{\Psis}{\Psi^{\mathbf{1}}}
\newcommand{\Psif}{\Psi^{\mathbf{5}}}

\begin{document}

\begin{titlepage}
\begin{flushright}
DFPD-2015/TH/11
\end{flushright}
\vskip1.8cm
\begin{center}
{\bf \LARGE{Beyond the Minimal Top Partner Decay} }
\vskip1cm 
{\bf Javi Serra}
\vskip 1cm
{\small \textit{Dipartimento di Fisica e Astronomia, Università di Padova \& INFN, Sezione di Padova, \\ Via Marzolo 8, I-35131 Padova, Italy}}
\vskip0.1cm
{\tt \href{mailto:jserra@pd.infn.it}{jserra@pd.infn.it}}
\end{center}
\vskip.5cm
\baselineskip 15pt
\begin{abstract}
Light top partners are the prime sign of naturalness in composite Higgs models. 
We explore here the possibility of non-standard top partner phenomenology. 
We show that even in the simplest extension of the minimal composite Higgs model, featuring an extra singlet pseudo Nambu-Goldstone boson, the branching ratios of the top partners into standard channels can be significantly altered, with no substantial change in the generated Higgs potential.
Together with the variety of possible final states from the decay of the pseudo-scalar singlet, this motivates more extensive analyses in the search for the top partners.
\end{abstract}
\date{January, 2015}
\end{titlepage}

\setcounter{equation}{0}
\setcounter{footnote}{0}


\section{Introduction} \label{intro}

Composite Higgs models aim at solving the electroweak hierarchy problem by postulating a new strongly interacting sector that dynamically generates the Higgs field. 
This emerges as a pseudo Nambu-Goldstone boson (pNGB), which explains why it is parametrically lighter than any typical composite resonance.
Ultimately, the Higgs is screened from high energy scales on account of its composite nature \cite{Kaplan:1983fs,Schmaltz:2005ky,Contino:2010rs,Bellazzini:2014yua}.

In this class of scenarios, a set of vector-like composite fermions linked to the top quark is responsible for keeping the Higgs potential under control \cite{ArkaniHamed:2002qx,Agashe:2004rs}. 
As long as the mass of the top partners is below the $\TeV$, the electroweak scale and the Higgs mass can be reproduced without significant fine tuning. 
This follows from a simple estimate, based on power counting and selection rules, of the size of the Higgs potential $V \simeq - \mu^2 |H|^2 + \lambda |H|^4$ generated by loops of the top and its partners,
\beq
\Delta \mu^2 \sim \frac{3 y_t^2}{8 \pi^2} m_T^2 \approx (90 \GeV)^2 \left( \frac{m_T}{500 \GeV} \right)^2 \ , \quad \Delta \lambda \sim \frac{3 y_t^2}{4 \pi^2} g_T^2 \approx 0.13 \left( \frac{g_T}{2} \right)^2 \ .
\label{eq:Htopmasslambda}
\eeq
The top Yukawa coupling $y_t$ is the largest coupling in the Standard Model (SM) that explicitly breaks the global shift symmetry protecting the Higgs. The top partner mass $m_T$ controls the size of the potential, while the top partner coupling, defined as $g_T \equiv m_T/f$, where $\fpi$ is the compositeness scale of the Higgs, determines the physical Higgs mass once the electroweak symmetry is broken. 
These estimates, verified in explicit constructions, point towards light and weakly coupled top partners saturating the radiatively generated Higgs potential \cite{Matsedonskyi:2012ym,Redi:2012ha,Pomarol:2012qf,Pappadopulo:2013vca}. 

The absence so far of any evidence of Higgs compositeness, in electroweak precision tests or Higgs couplings measurements, has pushed the scale $\fpi$ to somewhat unnatural values $f \gtrsim 600 \GeV$ \cite{Falkowski:2013dza,Grojean:2013qca}, and driven with it these models into the $\lesssim 10\%$ fine-tuned territory.
Besides, the ATLAS and CMS collaborations have also directly searched, without success, for the top partners potentially produced during the first LHC run \cite{Chatrchyan:2013wfa,Aad:2015kqa}. 
The lower bounds placed on their masses, $m_T \gtrsim 800 \GeV$, have started to build up the tension with naturalness. 
With the increase in energy and luminosity that will come with the second run of the LHC, the mass reach of direct searches will be substantially higher. 
Such an upgrade will provide an excellent opportunity for uncovering the symmetry mechanism protecting the Higgs potential and the agents implementing it, but it will also become a crucial test of the idea, given its present degree of tuning. 
In this regard, it is very important to understand the level of model dependence involved in the actual experimental searches of top partners.
Such searches are mainly based on pair production through QCD interactions (and seldom on single production via electroweak interactions), and decays to $W^\pm$, $Z$, or $h$, plus a top or a bottom quark.
However, there exist models, implementing the twin Higgs mechanism, in which the Higgs potential is controlled by top partners that are neutral under the SM gauge group, in particular under $\SU(3)_C$ color \cite{Chacko:2005pe}.
This possibility, although theoretically challenging, provides a proof of principle for natural theories with no direct signals at the LHC, at least of the standard kind. 
Another, more modest, approach towards unusual phenomenology for the top partners regards non-standard decay channels \cite{Kearney:2013oia,Kearney:2013cca}.
These could proceed via new light states, a natural option being other pNGBs. 
In fact, given our ignorance about the UV degrees of freedom participating of the strong dynamics, the appearance of extra light scalars in the IR is a well-motivated possibility. 
The mass of these extra scalars could receive contributions, along with the Higgs, from top loops, in which case 
\beq
\Delta m^2 \sim \epsilon \frac{3 y_t^2}{8 \pi^2} m_T^2 \ .
\label{eq:etatopmass}
\eeq
This kind of contributions are generically below the top partner masses $m_T$, given the implicit assumption that the couplings (in the case above $y_t$) that explicitly break the corresponding shift symmetry, are small perturbations.
Besides, extra parameters, denoted above by $\epsilon$, must always be kept in mind, to account for the different selection rules associated with the extra shift symmetry. 
These could actually render the entire top contribution vanishing, and the extra scalars naturally very light, as much as allowed by experimental searches.

In this work we study the feasibility of non-minimal top partner decays within a composite Higgs model featuring a single extra pNGB, in addition to the Higgs complex doublet.
The Next to Minimal Composite Higgs model (NMCHM) is based on a global $\SO(6)$ symmetry spontaneously broken to $\SO(5)$ \cite{Gripaios:2009pe}.\footnote{This model can be realized as a theory with four flavors of strongly interacting technifermions in a pseudo-real representation of the confining gauge group \cite{Galloway:2010bp,Barnard:2013zea}. \label{foot1}}
The extra light scalar $\eta$ is a singlet under the SM gauge symmetries, and we further take it to be a $\cp$-odd state. 
We will show that in this scenario a subset of the top partners can have a significant branching ratio into the pseudo-scalar singlet and a top quark, becoming even the dominant one under some circumstances.
This comes about without affecting the level of tuning required to reproduce the Higgs potential.
By focussing on two specific examples we will show how this can be possible. 
On the one hand, new sources of explicit breaking of the global symmetries can be introduced that, while giving rise to a dominant coupling of the top and its partner to $\eta$, do not directly break the shift symmetry protecting the Higgs. 
In the case these extra interactions do contribute to the Higgs potential, they do it in such a way as to reduce the overall contribution. 
On the other hand, the extended global symmetry structure predicts additional top partners that decay exclusively to the singlet. 
The phenomenology of $\eta$ is mainly dictated, as that of the Higgs, by considerations regarding the symmetries of the low energy effective theory. 
Depending on those symmetries and their breaking by the interactions with the SM fields, the singlet can present a varied pattern of decay channels, and therefore also the final products of the decays of the top partners can be variable.
Moreover, given that the compositeness scale of $\eta$ is the same as that of the Higgs, the phenomenology of the singlet is mainly controlled by dimension five operators suppressed by $\fpi$, which are not usually considered in collider studies of this type of scalars. 
The extra decay channel of the top partners to $\eta$, along with the diversity of decays of such a scalar, motivates extended searches for both particles. 

This paper is organized as follows. In section \ref{bmchm} we briefly review the symmetry structure of the NMCHM. The dependence of the Higgs and singlet potential on the top partners is presented in section \ref{potential} for two simple models. The phenomenology of the top partners is discussed in section \ref{pheno}, while that of the singlet pseudo-scalar can be found in section \ref{phenosinglet}. We conclude in section \ref{conclusions}.


\section{Beyond the Minimal Composite Higgs Model} \label{bmchm}

The Higgs complex doublet and an extra singlet $\eta$ arise as the NGBs of the spontaneous symmetry breaking $\SO(6)/\SO(5) \cong \SU(4)/\Sp(4)$ \cite{Gripaios:2009pe}. 
This coset indeed contains five scalar degrees of freedom, transforming in the $\mathbf{5}$ representation of $\SO(5)$.
This decomposes as a $\mathbf{1} + \mathbf{4} = (\mathbf{1},\mathbf{1}) + (\mathbf{2},\mathbf{2})$ of the custodial symmetry $\SO(4) \cong \SU(2)_L \times \SU(2)_R$. 
The associated Goldstone matrix, $U(\Pi) = \exp \left( i \sqrt{2} \Pi_{\ahat}(x) T^{\ahat} \right)$, can be conveniently written as 
\beq
U(\Pi) = 
\begin{pmatrix}
   1_{3 \times 3}  &  &  &  \\
       & 1 - \frac{h^2}{1+\sqrt{1-h^2-\eta^2}} & -\frac{h \eta}{1+\sqrt{1-h^2-\eta^2}} & h \\
      & -\frac{h \eta}{1+\sqrt{1-h^2-\eta^2}} & 1 - \frac{\eta^2}{1+\sqrt{1-h^2-\eta^2}} & \eta \\
       & -h & -\eta & \sqrt{1-h^2-\eta^2} \\  
\end{pmatrix}
\label{eq:umatrix} \ ,
\eeq
where we have eliminated the three NGBs eventually eaten by the $W^\pm$ and the $Z$. 
The Goldstone matrix transforms as $U(\Pi) \to g U(\Pi) \hat h^\dagger(\Pi,g)$, with $g$ a global $\SO(6)$ transformation and $\hat h$ a local (dependent on $\Pi(x)$) $\SO(5)$ transformation. 
When constructing the effective Lagrangians for the NGBs, we will often make use of a projector into the broken directions 
$\Sigma_0 = \begin{pmatrix}
 0  & 0  & 0  & 0  & 0  & 1      
\end{pmatrix}^T$. 
With it we can define the Goldstone multiplet 
$\Sigma = U(\Pi) \Sigma_0 = \begin{pmatrix}
 0  & 0  & 0  & h  & \eta  & \sqrt{1-h^2-\eta^2}      
\end{pmatrix}^T$.

The kinetic term for the NGBs is given by the leading invariant term in derivates, $\order{\partial^2}$,
\bea
\frac{\fpi^2}{4} d_\mu^{\ahat} d^\mu_{\ahat} = \frac{\fpi^2}{2} (D_\mu \Sigma)^\dagger (D^\mu \Sigma) 
\!\!\! &=& \!\!\! 
\frac{1}{2} (\partial_\mu h)^2 + \frac{1}{2} (\partial_\mu \eta)^2 + \frac{1}{2} \frac{\left( h\partial_\mu h + \eta \partial_\mu \eta \right)^2}{\fpi^2 -h^2 -\eta^2} \nn \\
&& + \frac{g^2}{4} h^2 \left( W_\mu^+ W^{\mu-} + \frac{1}{2 \cos^2 \theta_W} Z_\mu Z^\mu \right) \ ,
\label{eq:kinetic}
\eea
where we have given dimensions to the NGBs: $h \to h/\fpi$, $\eta \to \eta/\fpi$. 
The object $d_\mu$ is defined as $d_\mu^{\ahat} \equiv -i \, \Tr[T^{\ahat} U^\dagger D_\mu U]$, and it is one of the basic building blocks of our effective Lagrangians (see appendix~\ref{ccwz} for more details). 
As in the SM, once $h$ gets a vacuum expectation value (VEV), $\vev{h} = v \approx 246 \GeV$, the weak gauge bosons $W^\pm$ and $Z$ become massive. 
It is of phenomenological relevance that $h$ has extra derivative self-interactions and interactions with $\eta$. 
The first implies that after electroweak symmetry breaking (EWSB), the kinetic term of the Higgs receives an extra positive contribution of order $v^2/\fpi^2$, which has the net effect of suppressing all of the Higgs interactions.
The second gives rise, if kinematically allowed, to a non-standard Higgs decay to two $\eta$'s controlled by $1/\fpi$.

As far as the interactions in \Eq{eq:kinetic} are concerned, the scalar singlet can either be $\cp$-even or -odd. 
Actually, the Lagrangian \Eq{eq:kinetic} is invariant under a set of discrete $Z_2$ transformations that act individually on each of the NGBs as $\Pi_{\ahat} \to - \Pi_{\ahat}$, as well as under the spacetime parity $P_0$: $x \to - x \, , t \to t$, $\Pi_{\ahat} \to \Pi_{\ahat}$. 
We will be particularly interested in the combination $CP = C_6 P_6 P_0$, which defines the $\cp$ symmetry of the NGBs in $\SO(6)/\SO(5)$: $h \to h$ and $\eta \to - \eta$. 
The automorphism $C_6$ is identified with charge conjugation, while $P_6$ corresponds to the grading of the algebra, under which all the unbroken generators remain unchanged $T^{a} \to +T^{a} \ \forall \, a$, while the broken generators change sign, $T^{\ahat} \to -T^{\ahat} \ \forall \, \ahat$.\footnote{In the basis we have used to write the Goldstone matrix in \Eq{eq:umatrix}, these discrete transformations are given by $\mathcal{C}_6 = \diag(-1, 1, -1, 1, -1, -1)$ and $\mathcal{P}_{6} = \diag(1, 1, 1, 1, 1, -1)$. Notice that the parity $P_{6}$ is actually an outer automorphism, not contained in $\SO(6)$, and generically it should not be respected by higher order terms in the Lagrangian expansion in derivatives.}
In this work we will assume that the strong sector respects $\cp$, and that it remains unbroken to a high degree of approximation by the interactions with the SM fields (keeping in mind the amount of $\cp$ violation needed to reproduce the SM). 
This assumption is in fact necessary to avoid too large contributions to $\cp$-violating observables. 
Furthermore, the $\SO(6)/\SO(5)$ coset admits a Wess-Zumino-Witten (WZW) term \cite{Witten:1983tw}, arising at the next to leading order in derivatives, $\order{\partial^4}$, that respects $\cp$. 
This term could play an important role in the phenomenology of $\eta$, since it gives rise, at leading order in $\fpi$, to the interactions:
\beq
\frac{\eta}{\fpi} \frac{\epsilon^{\mu \nu \rho \sigma}}{48 \pi^2} \sum_{a=a_C, a_L,Y} n_a g_a^2 F^a_{\mu \nu} F^a_{\rho \sigma} \ ,
\label{eq:anomaly}
\eeq
where $F^a_{\mu \nu}$ are the field strengths of the SM $\SU(3)_C \times \SU(2)_L \times \UU(1)_Y$ gauge group, $g_a$ the corresponding gauge couplings, and $n_a$ the anomaly coefficients, which carry information about the underlying UV structure of the theory. 
In particular, given the $\SU(4) \times \SU(3)_C$ global symmetry structure of the strong sector under consideration, $n_g = 0$ and $n_W = - n_B$.


\section{Potential and Top Partner Masses} \label{potential}

The potential for the pNGBs depends on how the associated shift symmetries are explicitly broken. 
The SM already contains relevant symmetry breaking parameters: the top Yukawa coupling, and the $\SU(2)_L$ gauge coupling. 
In order to understand how the global symmetries are broken, we need to specify how the top quark, $q_L$ and $t_R$, and the gauge bosons $W^a$ are coupled to the strong sector. 
The latter is fixed by gauge invariance, that is gauge fields couple to the strong sector's associated conserved currents $\Lag \supset g W^a_\mu \mathcal{J}^{\mu \, a}$. 
The former depends on how the top Yukawa coupling is generated. 
We will be assuming that the top quark couples via mixing with composite fermionic operators in the UV, \ie partial compositeness: $\Lag \supset \lambda_L \bar q_L \Op_{q} + \lambda_R \bar \Op_t t_R + h.c.$ \cite{Kaplan:1991dc}. 
Given that both the operators $\Op_{q,t}$ and the current $\mathcal{J}^{\mu \, a}$ are part of entire representations of $\SO(6)$ (the first to be specified, and the second in the adjoint $\mathbf{15}$), while the SM fields do not fill complete $\SO(6)$ multiplets, these interactions break explicitly the global symmetries. 
Notice that in partial compositeness the breaking introduced by the top Yukawa is a consequence of the combined breaking introduced by $\lambda_L$ and $\lambda_R$, since $y_t \propto \lambda_L \lambda_R$. 
We should observe as well that, in order to reproduce the correct hypercharges of the $\Op_{q,t}$ components mixing with $q_L$ and $t_R$, an extra unbroken $\UU(1)_X$ global symmetry must be introduced. 
The hypercharge of the states of the strong sector is then given by $Y = T_R^3 + X$, where $T_R^3$ is the $\UU(1)$ generator inside $\SU(2)_R$. 
While the pNGBs do not carry $X$ charge, we will assume that the fermionic operators $\Op_{q,t}$ (and their associated resonances, \ie the top partners) have $X = 2/3$. 

The potential induced by loops of the $\SU(2)_L$ gauge bosons can be derived once we properly identify which generators of $\SO(6)$ are associated to the $\SU(2)_L$ current (see appendix~\ref{ccwz} for an explicit expression).
Such generators, times the weak gauge coupling $g$, can then be viewed as spurionic fields, from which $\SO(6)$ invariants can be constructed. 
At leading order, $\order{g^2}$, they lead to the potential: 
\beq
V_{g^2} = c_g f^2 \sum_{a=a_L} \Sigma^T (g T_L^{a}) (g T_L^{a}) \Sigma = c_g \frac{3}{4} g^2 h^2 \ ,
\label{eq:Wloop}
\eeq
where we recall that $\Sigma = U \Sigma_0$.
Notice that only a mass term for the Higgs is generated, but there is no potential for the $\eta$. 
This is a consequence of the fact that $\eta$ is a singlet under $\SU(2)_L$, thus the gauging of $\SU(2)_L$ does not break the $\UU(1)_\eta$ shift symmetry protecting the singlet. 
We can estimate the size of the coefficient as $c_g \sim 3 m_\rho^2 / 32 \pi^2$, where $m_\rho$ is the mass of the vector resonances cutting off the loop of $W$'s. 
For $m_\rho = 2.5 \TeV$, this contribution implies a moderate tuning of $\approx 20 \%$ on the Higgs mass term.

In a similar fashion we can identify the contributions to the potential from loops of $q_L$ and/or $t_R$. 
For this we need to specify the transformation properties of the operators $\Op_{q,t}$ the top couples to, and then the actual interactions $\lambda \bar t \Op$ will only be restricted by the requirement that they should respect the SM gauge symmetries. 
It will be convenient to use the embedding fields $Q_L = b_L \, \upsilon_{b_L}  + t_L \, \upsilon_{t_L}$ and $T_R = t_R \, \upsilon_{t_R}$, in order to write the couplings of the top to the composite fermionic operators as $\Lag \supset \lambda_L (\bar Q_L)_I \Op_{q}^I + \lambda_R (\bar \Op_t)_I (T_R)^I$, where the index $I$ runs over $\SO(6)$ components.
Then $\lambda_L \upsilon_{b_L}$, $\lambda_L \upsilon_{t_L}$, and $\lambda_R \upsilon_{t_R}$ can be treated as spurions from which we can compute the potential. 

We will be considering two different sets of representations for the operators $\Op_{q,t}$. 
For the first we will assume that both $\Op_{q,t}$ transform in vector $\mathbf{6}$ representation of $\SO(6)$, with $X=2/3$.\footnote{This is the extension to $\SO(6)/\SO(5)$ of the minimal $\SO(5)/\SO(4)$ model with the $L$- and $R$-handed top embedded in the vector $\mathbf{5}$ representation of $\SO(5)$ \cite{Contino:2006qr}.}
In that case the embeddings of $q_L$ and $t_R$ are given by
\beq
\mathbf{6}_L: \quad \upsilon_{b_L} = \frac{1}{\sqrt{2}} 
\begin{pmatrix}
 i  & + 1  & 0  & 0  & 0  & 0      
\end{pmatrix}^T
\, , \,
\upsilon_{t_L} = 
\frac{1}{\sqrt{2}}
\begin{pmatrix}
 0  & 0  & i  & -1  & 0  & 0      
\end{pmatrix}^T \ .
\label{eq:embed6L} 
\eeq
\beq
\mathbf{6}_R: \quad \upsilon_{t_R} =
\begin{pmatrix}
 0  & 0  & 0  & 0  & i \, \gamma  & 1      
\end{pmatrix}^T \ .
\label{eq:embed6R}
\eeq
Notice that the embedding in \Eq{eq:embed6R} implies that $t_R$ couples to two different components of $\Op_t$, with relative strengths set by $\gamma$. 
This parameter can be taken to be real and positive without loss of generality. 
The couplings of $q_L$ and $t_R$ specified by the above embeddings lead to the potential, at leading order in the symmetry breaking couplings, $\order{\lambda_L^2}$ and $\order{\lambda_R^2}$,
\bea
\label{eq:VL2}
V_{\lambda_L^2}
\!\!\! &=& \!\!\!
c_L f^2 \sum_{\alpha=t_L,b_L} (\Sigma^T \lambda_L \upsilon_\alpha) (\lambda_L \upsilon_\alpha^\dagger \Sigma) = c_L \lambda_L^2 \, \frac{h^2}{2} \ . \\
\label{eq:VR2}
V_{\lambda_R^2}
\!\!\! &=& \!\!\!
c_R f^2 (\Sigma^T \lambda_R \upsilon_{t_R}) (\lambda_R \upsilon_{t_R}^\dagger \Sigma) = c_R \lambda_R^2 \, \left[ f^2 - h^2 + (\gamma^2 - 1) \eta^2 \right] \ .
\eea
We would like to point out several important aspects of \Eqs{eq:VL2} and (\ref{eq:VR2}). 
First, there is no contribution to any $h$ dependent term from $\gamma \lambda_R$, given that this coupling does not break the Goldstone-Higgs shift symmetry. 
Second, there is no contribution to any $\eta$ dependent term when $\gamma \to 1$, given that in this limit there is no breaking of the Goldstone-singlet shift symmetry. 
In particular, the interaction of $q_L$ with the strong sector does not break the $\UU(1)_\eta$, only those of $t_R$ do for $\gamma \neq 1$. 
This is understood by observing that one can formally assign to $q_L$ a definite $\UU(1)_\eta$ charge, $T_\eta Q^{\mathbf{6}}_L = 0 \, Q^{\mathbf{6}}_L$, and likewise for $t_R$, $T_\eta T^{\mathbf{6}}_R = -1/\sqrt{2} \, T^{\mathbf{6}}_R$, provided $\gamma = 1$. 
A simple estimate leads to the coefficients $c_{L,R} \sim 6 m_\Psi^2 / 16 \pi^2$, where $m_\Psi$ is the mass scale at which the top loop is cut off. 
We will compute below in a specific example the actual dependence of $c_{L,R}$ on the top partners masses. 
From the embeddings of $q_L$ and $t_R$ in \Eqs{eq:embed6L} and (\ref{eq:embed6R}) we can also derive the top Yukawa coupling:
\beq
y_t f (\bar Q^{\mathbf{6}}_L \Sigma) (\Sigma^T T^{\mathbf{6}}_R) + h.c. = - \frac{y_t}{\sqrt{2}} \bar t_L h t_R \left( \sqrt{1 -\frac{h^2}{\fpi^2}-\frac{\eta^2}{\fpi^2}} + i \gamma \frac{\eta}{f} \right) + h.c. \ .
\label{eq:topYuk}
\eeq
Finally, we must notice that a Higgs quartic term is not generated at leading order in $\lambda_L$ or $\lambda_R$. 
This must therefore come from subleading $\order{\lambda^4}$ terms. 
One of such terms (the one that actually descends from the top Yukawa coupling above), reads,
\beq
V_{\lambda_L^2\lambda_R^2} = c_{LR} f^4 \sum_{\alpha=t_L,b_L} | (\Sigma^T \lambda_R \upsilon_{t_R}) (\lambda_L \upsilon_\alpha^\dagger \Sigma)|^2 = 
c_{LR} \lambda_L^2\lambda_R^2 \, \frac{h^2}{2} \left[ f^2 - h^2 + (\gamma^2 - 1) \eta^2 \right] \ ,
\label{eq:VL2R2}
\eeq
with the estimate $c_{LR} \sim 6/16\pi^2$. 
The fact that any term in the potential involving the singlet vanishes in the limit $\gamma \to 1$ remains true at any order in the breaking parameter $\lambda_R$. 

We will also be considering the alternative option in which the operator $\Op_{q}$ transforms in the $\mathbf{20'}$ (symmetric and traceless component of $\mathbf{6} \times \mathbf{6}$) representation of $\SO(6)$, while $\Op_{t}$ is just a singlet $\mathbf{1}$, both with $X = 2/3$. 
Since in this case $t_R$ has a trivial embedding, $\lambda_R$ does not give rise to any explicit breaking.
We only need to specify the embedding of $q_L$, in a symmetric traceless tensor:
\beq
\mathbf{20'}_L: \quad \hat \upsilon_{b_L} = \frac{1}{\sqrt{2}} 
\begin{pmatrix}
 0_{4 \times 4}  & -i \hat \gamma \upsilon_{b_L}  & \upsilon_{b_L} \\
 - i \hat \gamma \upsilon_{b_L}^T  & 0  & 0 \\
   \upsilon_{b_L}^T  & 0  & 0
\end{pmatrix}
\, , \,
\hat \upsilon_{t_L} = \frac{1}{\sqrt{2}}
\begin{pmatrix}
 0_{4 \times 4}  & - i \hat \gamma \upsilon_{t_L}  & \upsilon_{t_L} \\
 - i \hat \gamma \upsilon_{t_L}^T  & 0  & 0 \\
   \upsilon_{t_L}^T  & 0  & 0
\end{pmatrix}
\label{eq:embed20L} \ ,
\eeq
where $\upsilon_{b_L, t_L}$ have been given in \Eq{eq:embed6L}, and $\hat \gamma \in \mathbb{R}^+$, consistently with $\cp$ symmetry. 
This embedding of $q_L$ gives rise, at leading order in $\lambda_L$, to two different invariants in the potential:
\bea
\label{eq:VL2a}
V^{(1)}_{\lambda_L^2}
\!\!\! &=& \!\!\!
c^{(1)}_L f^2 \sum_{\alpha=t_L,b_L} \Sigma^T (\lambda_L \hat \upsilon_\alpha) (\lambda_L \hat \upsilon_\alpha)^\dagger \Sigma = c^{(1)}_L \lambda_L^2 \, \left[ 1 + (\hat \gamma^2 - 3) \frac{h^2}{4} + (\hat \gamma^2 - 1) \eta^2 \right] \ , \\
\label{eq:VL2b}
V^{(2)}_{\lambda_L^2}
\!\!\! &=& \!\!\!
c^{(2)}_L f^2 \sum_{\alpha=t_L,b_L} (\Sigma^T \lambda_L \hat \upsilon_\alpha \Sigma) (\Sigma^T \lambda_L \hat \upsilon_\alpha \Sigma)^\dagger = c^{(2)}_L \lambda_L^2 \, h^2 \left[ 1 - h^2 + (\hat \gamma^2 - 1) \eta^2 \right] \ .
\eea
The interactions of $q_L$ now break the $\UU(1)_\eta$ shift symmetry, whenever $\hat \gamma \neq 1$. 
Conversely, the singlet becomes massless in the limit $\hat \gamma \to 1$. 
The Higgs potential receives contributions from $\lambda_L$ and also from $\hat \gamma \lambda_L$. 
However, moderate values of $\hat \gamma$ tend to reduce the Higgs mass term. 
This will become clear in the specific example presented below. 
Contrary to the previous embeddings, a Higgs quartic is generated at leading order in the breakings. 
Therefore it is not necessary to involve subleading terms to reproduce the Higgs potential. 
Finally, the top Yukawa coupling is given by the $\SO(6)$ invariant
\beq
\frac{y_t}{\sqrt{2}} f (\Sigma^T \bar Q^{\mathbf{20'}}_L \Sigma) t_R + h.c. = - \frac{y_t}{\sqrt{2}} \bar t_L h t_R \left( \sqrt{1 -\frac{h^2}{\fpi^2}-\frac{\eta^2}{\fpi^2}} + i \hat \gamma \frac{\eta}{f} \right) + h.c. \ .
\label{eq:topYuk20}
\eeq

In the next sections we will present simple realizations of the two cases considered above, where the fermionic operators $\Op_{q,t}$ are interpreted as light composite resonances, below the mass of the cutoff $\Lambda \lesssim 4 \pi f$. 
We will assume that these top partners saturate the Higgs potential, in order to gain a qualitative and somewhat quantitative understanding of their role in reproducing the electroweak VEV and the Higgs mass.

We should also keep in mind that other possible sources of explicit breaking beyond those associated to the SM could be present. 
For instance, in the present scenario a plausible source of breaking could be given by
\beq
- c_M M^T \Sigma = - c_M m \sqrt{1-h^2-\eta^2} \ ,
\label{eq:technimass}
\eeq
where $M \equiv m \Sigma_0$ and $c_M \sim 4 \pi f^3$.\footnote{This contribution could originate from a non-vanishing mass term for the technifermions, see foonote \ref{foot1}. 
In the estimate of $c_M$ we have assumed the free field scaling for the technifermion bilinear.} 
If the mass $m$ is a relevant perturbation at the compositeness scale, the term above could have a significant impact in the pNGB potential \cite{Galloway:2010bp}.

The last comment we want to make regards other possible sources of explicit symmetry breaking that only affect the singlet. 
Consider for instance the couplings of the $R$-handed bottom to the strong sector. 
If we assume that $b_R$ is embedded in a $\mathbf{6}$ of $\SO(6)$, the subsequent contributions to the pNGB potential will be similar to those of $t_R$ in \Eq{eq:VR2}, with $\lambda_R \to \lambda_R^{b}$ and $\gamma \to \gamma_b$. 
Therefore, if $\gamma_b \gg 1$ an important contribution to the mass of the singlet could be generated from bottom loops, while still reproducing the small bottom Yukawa $y_b \sim \lambda_R^{b} \lambda_L^{b}$, and without contributing significantly to the Higgs potential (notice in particular that the spurion $\gamma_b \lambda_R^b$ has different $\cp$ quantum numbers than $y_b$). 
This is just one possibility that reflects the fact that the potential for $\eta$ is subject to more model dependencies than that of the Higgs.


\subsection{Toy Model $\mathbf{6}_L+\mathbf{6}_R$} \label{model1}

With a simple effective Lagrangian containing the top partners, we can understand how their masses fix the coefficients $c_{L,R}$ and $c_{LR}$ in \Eqs{eq:VL2}, (\ref{eq:VR2}), and (\ref{eq:VL2R2}). 
To this aim, we introduce a complete multiplet of massive top partners in the vector $\mathbf{6}$ representation of $\SO(6)$,
\beq
\Psi_{L,R} = \begin{pmatrix}
      \Psif    \\
      \Psis      
\end{pmatrix}_{L,R}
\, , \quad
\Psif_{L,R} = \frac{1}{\sqrt{2}} \begin{pmatrix}
      i (B-X_{5/3})    \\
      B+X_{5/3}    \\
      i (X_{2/3}+T)    \\
      X_{2/3}-T    \\
      \sqrt{2} \, i \, T'    \\     
\end{pmatrix}_{L,R} \ .
\label{eq:psi6}
\eeq
As the notation suggests, $\Psi$ decomposes under $\SO(5)$ as a 5-plet $\Psif$ and a singlet $\Psis$. 
The Lagrangian for the top sector then reads,
\beq
-\Lag_\Psi = \lambda_\Psi f \bar \Psi_L \Psi_R - y_\Psi f (\bar \Psi_L \Sigma) (\Sigma^T \Psi_R) + \lambda_L f \bar Q^{\mathbf{6}}_L \Psi_R + \lambda_R f \bar \Psi_L T^{\mathbf{6}}_R + h.c. \ ,
\label{eq:Ltop6a}
\eeq
where the embeddings $Q^{\mathbf{6}}_L$ and $T^{\mathbf{6}}_R$ have been identified in \Eqs{eq:embed6L} and (\ref{eq:embed6R}). 
Such a Lagrangian is often found in 2-site descriptions of composite Higgs models \cite{Matsedonskyi:2012ym}. 
Its symmetry features are clear once we perform a $\SO(6)$ rotation on $\Psi_{L,R}$ that eliminates the NGB dependence in the $y_\Psi$ term, moving it to the mixing terms $\lambda_{L,R}$,
\beq
-\tilde \Lag_\Psi = M_5 \bar \Psi^{\mathbf{5}}_L \Psif_R + M_1 \bar \Psi^{\mathbf{1}}_L \Psis_R + \lambda_L f \bar Q^{\mathbf{6}}_L U \Psi_R + \lambda_R f \bar \Psi_L U^\dagger T^{\mathbf{6}}_R + h.c. \ ,
\label{eq:Ltop6}
\eeq
where $M_5 = \lambda_\Psi f$ and $M_1 = (\lambda_\Psi - y_\Psi) f$ (notice that the masses of the 5-plet and the singlet are independent).
The collective pattern of $\SO(6)$ symmetry breaking is now apparent. 
Both $y_\Psi = (M_5 - M_1)/\fpi$ and $\lambda_L$ or $\lambda_R$ are needed in order to generate a non-trivial potential for the NGBs. 
This also implies that any one-loop contribution to the potential will be at most logarithmically divergent within this simple model.

The top partners masses, at leading order in $\lambda_{L,R}$ and neglecting EWSB effects, which are suppressed by $v^2/f^2$, are given by
\bea
m_{X_{5/3}} = m_{X_{2/3}} = M_5
\ , && 
m_B \simeq m_T \simeq \sqrt{M_5 ^2+ (\lambda_L f)^2}
\ , \nn \\
 m_{T'} \simeq \sqrt{M_5^2 + (\gamma \lambda_R f)^2}
\ , &&
m_{\Psis} \simeq \sqrt{M_1^2 + (\lambda_R f)^2} \ .
\label{eq:masses}
\eea
The top Yukawa coupling in \Eq{eq:topYuk}, arising through the mass-mixing of the elementary states $q_L $ and $t_R$ and the composite resonances in $\Psi$, is given by
\beq
y_t = y_\Psi \frac{\lambda_L f}{m_T} \frac{\lambda_R f}{m_{\Psis}} \frac{m_{X_{5/3}}}{m_{T'}}
\label{eq:Yukcoupling} \ .
\eeq
From the Lagrangian \Eq{eq:Ltop6a} it is clear why in order to generate a top Yukawa the couplings $y_\Psi$, $\lambda_L$, and $\lambda_R$ are needed. 
The first is the Yukawa-type coupling for the $\Psi$ fields, while the last two give rise to the necessary mixing angles $\lambda_L f/m_T$ and $\lambda_R f/m_{\Psis}$ for $q_L$ and $t_R$ respectively. 
Besides, the factor of $m_{X_{5/3}}/m_{T'}$ arises from the extra mixing of $t_R$ with $T'_R$. 
This extra factor favors large values of $M_5$ in order to reproduce the large top Yukawa. 

A standard computation of the Coleman-Weinberg potential yields the following result for the coefficients $c_L$ and $c_R$ in \Eqs{eq:VL2} and (\ref{eq:VR2}):
\beq
c_L = c_R = \frac{3}{8 \pi^2} \left[ M_1^2 \log \left( \frac{\Lambda^2}{M_1^2} \right) - M_5^2 \log \left( \frac{\Lambda^2}{M_5^2} \right) \right] \ .
\label{eq:c2}
\eeq
As expected, these are logarithmically divergent, the scale $\Lambda$ to be interpreted as the mass of a second layer of heavier fermionic resonances. 
Furthermore, $c_L$ and $c_R$ vanish in the limit $y_\Psi = M_5 - M_1 \to 0$, as we advanced after inspecting the symmetry properties of the top Lagrangian. 
The coefficient $c_{LR}$ in \Eq{eq:VL2R2} is instead finite at one loop, since it requires four insertions of the symmetry breaking couplings $\lambda_{L,R}$,
\beq
c_{LR} = \frac{3}{4 \pi^2} \frac{1}{M_5^2 - M_1^2} \left[ M_1^2 - M_5^2 + M_1 M_5 \log \left( \frac{M_5^2}{M_1^2} \right) \right] \ .
\label{eq:c4}
\eeq
Similar expressions are obtained for the terms arising at order $\lambda_L^4$ and $\lambda_R^4$. 
To understand under which conditions and with how much tuning the Higgs potential can be reproduced in this simple model, we must take into account that the top Yukawa coupling \Eq{eq:Yukcoupling} establishes a relation between the couplings $\lambda_{L,R}$ and the top partners masses. 
It follows then that the leading contributions to the Higgs mass term, expressed in terms of the mass of the top partners $\Psis$ and $X_{5/3}$, are 
\beq
\zeta (\Delta \mu^2)_L \simeq -\frac{1}{2 \zeta}(\Delta \mu^2)_R \simeq \mp \frac{3 y_t}{8 \pi^2} \frac{ m_{X_{5/3}} m_{\Psis} }{\fpi |m_{\Psis} \pm m_{X_{5/3}}|} \left[ m_{X_{5/3}}^2 \log \left( \frac{\Lambda^2}{m_{X_{5/3}}^2} \right) - m_{\Psis}^2 \log \left( \frac{\Lambda^2}{m_{\Psis}^2} \right) \right] \ ,
\label{eq:mH2}
\eeq
where we have defined $\zeta \equiv \lambda_R/\lambda_L$. 
These contributions scale as $\Delta \mu^2 \sim (m_\Psi/f)^3 f^2$, that is with the third power of the top partner's mass.
The two contributions are equal in size but opposite in sign when $\zeta = 1/\sqrt{2}$. 
This can be traced back to the fact that both $q_L$ and $t_R$ have been embedded in the same $\mathbf{6}$ representation of $\SO(6)$. 
At $\order{\lambda_{L,R}^2}$ there is no effect of a non-vanishing $\gamma$.
This arises at the next to leading order, primarily from the dependence introduced through the Yukawa of the top, and it is then suppressed by $(\gamma \lambda_R f/M_5)^2$. 
Whenever this ratio is small, we can approximate
\beq
(\Delta \mu^2)_{\gamma^2} \simeq \gamma^2 y_t \zeta \frac{f \min(m_{X_{5/3}},m_{\Psis})}{2 m_{X_{5/3}}^2} (\Delta \mu^2)_R \ .
\label{eq:mH2gamma}
\eeq
This contribution increases the Higgs mass term in the region of small $m_{X_{5/3}}$. 
We should notice though that for $\gamma \neq 0$, there is a lower theoretical bound on $m_{X_{5/3}}$, $(m_{X_{5/3}})_{min} = \gamma y_t f$, which arises from the requirement to reproduce the large top Yukawa. 
This is the main effect of a non-vanishing $\gamma$ in what regards the Higgs potential. 

Given the current bounds on $f$ and the masses of the top partners, the contributions in \Eq{eq:mH2} must be finely cancelled in order to reproduce the correct Higgs mass term.
Since in this simple model $c_L = c_R$, the cancellation can be achieved by adjusting $\zeta \simeq 1/\sqrt{2}$, and the level of tuning is higher the heavier are the top partners. 
The addition of the gauge contribution in \Eq{eq:Wloop} (generically positive) allows a departure from the relation $\lambda_R \simeq \lambda_L/\sqrt{2}$, how important depending on how heavy the composite vector resonances are. 
Likewise for an extra contribution from \Eq{eq:technimass} (also positive). 
In this regard, notice that this latter term can only play a role in the Higgs potential (given $c_M \sim 4 \pi f^3$) if $m/f \gtrsim y_t (m_\Psi/f)^3 / (4\pi)^3 \approx 10^{-3}$ for $m_\Psi/f = 2$.
In this simple model however there cannot be a large departure from $\zeta = 1/\sqrt{2}$, because in that case the Higgs quartic is not reproduced, see the discussion after \Eq{eq:quartic}.
One other possibility to tune down $\mu^2$ is to consider $M_1 < 0$. 
In that case the contributions in \Eq{eq:mH2} can be made small in the regime $m_{\Psis}^2 \simeq m_{X_{5/3}}^2$, while still reproducing the top Yukawa (recall that $y_t \sim |M_5 - M_1|$).
On top of this, the effect of a non-vanishing $\gamma$ in \Eq{eq:mH2gamma} is to disfavor the regions where $m_{X_{5/3}} \ll m_{\Psis}$, basically because of $(m_{X_{5/3}})_{min} \propto \gamma$. 
On the other hand, for $m_{X_{5/3}} \gg m_{\Psis}$, the dependence on $\gamma$ becomes small.
Therefore $\gamma \neq 0$ favors a light singlet top partner in this simple model. 
We will show in section \ref{pheno} that for $\gamma \gtrsim 1$ $\Psis$ decays predominantly to $\eta \, t$.

The masses of the top partners determine also the Higgs quartic coupling. 
The leading contribution, which arises at $\order{\lambda_{L,R}^4}$, takes a simple form in the limit $\zeta \to 1/\sqrt{2}$:
\beq
\lambda \simeq \frac{3 y_t^2}{4 \pi^2} \frac{m_{X_{5/3}}^2 m_{\Psis}^2}{f^2 (m_{X_{5/3}}^2 - m_{\Psis}^2)} \log \left( \frac{m_{X_{5/3}}^2}{m_{\Psis}^2}\right) \ .
\label{eq:quartic}
\eeq
This scales with the second power of the top partner's mass $\lambda \sim (m_\Psi/f)^2$.
It follows then that reproducing the lightness of the Higgs requires that one of the top partners, either the singlet or the 5-plet, is weakly coupled, $g_\Psi = (m_\Psi/f) \lesssim 2$.
Departures from the relation $\lambda_R = \lambda_L/\sqrt{2}$ increase the contribution of the top partners to the Higgs quartic, and therefore reproducing it requires even smaller $g_\Psi$. 
Taking into account in addition the theoretical lower bound on $m_{X_{5/3}}$ (for $\gamma \neq 0$), and the one present also for the mass of $\Psis$, $(m_{\Psis})_{min} = \zeta y_t f$, one finds that $\zeta \in (1/3 , 1)$, approximately. 

\begin{figure}[!t]
\begin{center}
\includegraphics[width=3in]{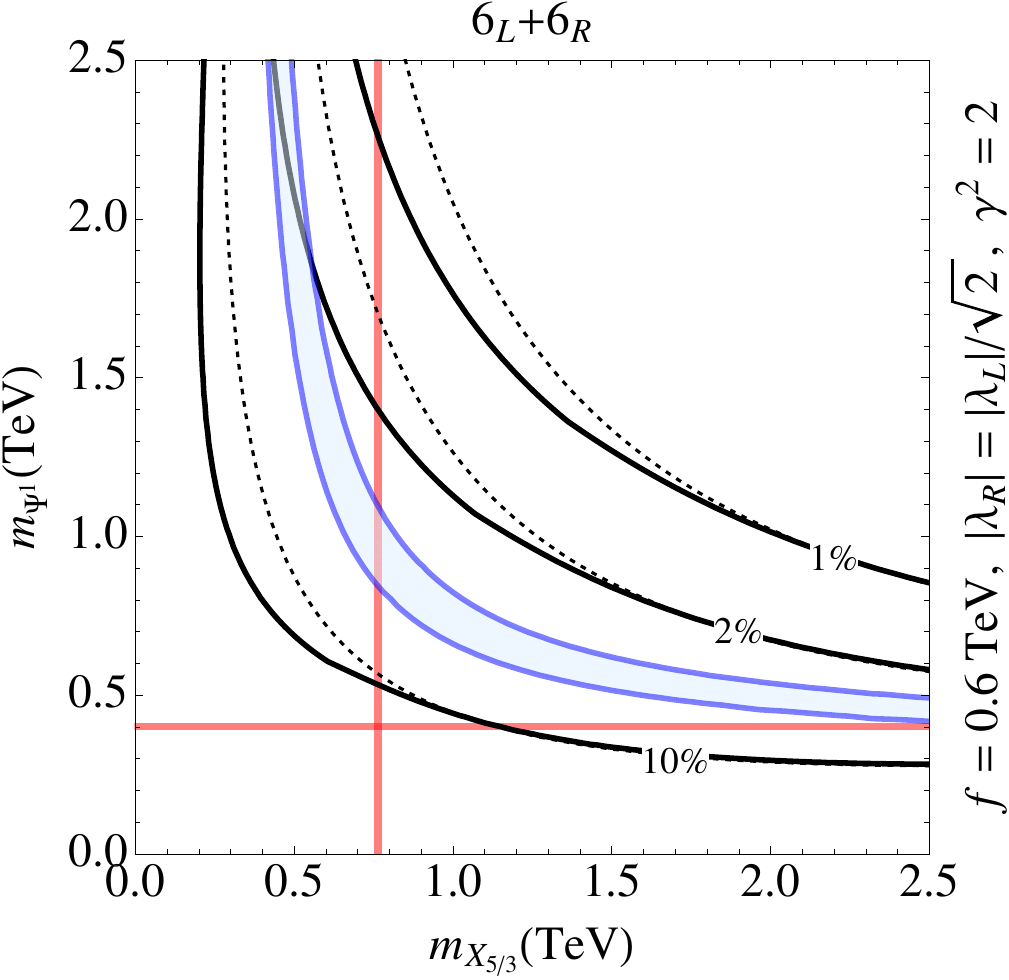}
\caption{Contour lines of tuning $\mu^2/(\Delta \mu^2)_R$ (solid black) and regions with Higgs quartic $0.11 \leqslant \lambda \leqslant 0.14$ (blue), in the plane of the top partners masses $m_{X_{5/3}}$ and $m_{\Psis}$, for $\gamma^2 = 2$, $\zeta = 1/\sqrt{2}$, and $f= 0.6 \TeV$. 
Contour lines of tuning for $\gamma = 0$ (dotted black) are also shown for comparison. 
The red lines delimitate the region (upper-right) where the top Yukawa can be reproduced. 
For $\gamma = 0$ the lower bound on $m_{X_{5/3}}$ goes to zero. 
Notice that for large $m_{X_{5/3}}$ there is little difference between the solid and dashed lines. 
We have taken $\Lambda = 4 \sqrt{m_{X_{5/3}} m_{\Psis}}$.}
\label{fig:tuningmass66}
\end{center}
\end{figure}

To illustrate the points discussed above, we show in Figure \ref{fig:tuningmass66} contour lines for the tuning in this model, as well as the region where the Higgs quartic is reproduced, in the plane ($m_X, m_{\Psis})$, and for a representative set of parameters. 
Let us stress that the numbers in this plot have been obtained under the approximations explained above, but its qualitative features properly reflect the effects of the top partners on the Higgs potential. 
We simply defined the tuning as the ratio of the largest contribution to $\mu^2$, which for the parameters taken in the plot corresponds to $(\Delta \mu^2)_R$, over its correct value $\mu^2 \approx (90 \GeV)^2$. 

The contribution of the top sector to the mass of the singlet is correlated with the degree of tuning to be enforced on the Higgs mass term,
\beq
(\Delta m_\eta^2)_R = (\gamma^2-1) (\Delta \mu^2)_R \approx \pm (320 \GeV)^2 \left( \frac{|\gamma^2-1|}{1} \right) \left( \frac{8\%}{\mu^2/(\Delta \mu^2)_R} \right) \ .
\label{eq:etamass}
\eeq
It is important to recall that in the limit $\gamma \to 1$ the singlet does not receive a potential from the top sector. 
It follows from \Eqs{eq:mH2} and (\ref{eq:etamass}) that if EWSB is driven by a positive $(\Delta \mu^2)_L$, then $(\Delta \mu^2)_R$ is negative, and $m_\eta^2$ is positive only for $\gamma < 1$. 
$\gamma > 1$ is only possible in this case, while keeping $m_\eta^2 > 0$, if some other contribution overcompensates \Eq{eq:etamass}. 
If such a contribution also adds to the Higgs potential, like \Eq{eq:technimass} \cite{Galloway:2010bp}, then it must be a leading one. 
This in turn requires $\gamma \sim 1$ in order not to increase the level of fine tuning. 
In contrast, if EWSB is due to $(\Delta \mu^2)_R > 0$, then $m_\eta^2 > 0$ for $\gamma > 1$, while for $\gamma < 1$ extra contributions are needed to keep $\eta$ from getting a VEV. 
Notice that we are focussing on $\vev{\eta} = 0$ to keep the Higgs from inheriting the properties of a pseudo-scalar singlet. 
In any case we should keep in mind that $m_\eta$ is exposed to large model dependencies. 


\subsection{Toy Model $\mathbf{20'}_L+\mathbf{1}_R$} \label{model2}

Here a $\mathbf{20'}$ multiplet of Dirac fermions $\Psi$ is coupled to $q_L$, breaking explicitly the $\SO(6)$ symmetry, while $t_R$ couples to a singlet of $\SO(6)$. 
Given that this latter mixing does not introduce any explicit breaking, we can actually dispense with the composite singlet and directly introduce a coupling of $t_R$ to the $\SO(5)$ singlet component of the $\mathbf{20'}$. 
Then the effective Lagrangian, in the field basis where the NGB dependence comes with the elementary-composite coupling $\lambda_L$, reads
\beq
-\tilde \Lag_\Psi = M_{14} \bar \Psi^{\mathbf{14}}_L \Psi^{\mathbf{14}}_R + M_5 \bar \Psi^{\mathbf{5}}_L \Psif_R + M_1 \bar \Psi^{\mathbf{1}}_L \Psis_R + \lambda_L f \Tr[\bar Q^{\mathbf{20'}}_L U \Psi_R U^T] + \lambda_R f \bar \Psi^{\mathbf{1}}_L t_R + h.c. \ ,
\label{eq:Ltop20p}
\eeq
where the embedding $Q^{\mathbf{20'}}_L$ has been given in \Eq{eq:embed20L}, and an explicit matrix form for $\Psi$ is given in appendix~\ref{ccwz}. 
Notice that $\Psi$ decomposes as a $\mathbf{1} + \mathbf{5} + \mathbf{14}$ of $\SO(5)$. 
From \Eq{eq:Ltop20p} one can understand that $\lambda_L$ is needed to generate a potential for the Higgs and $\eta$. 
After performing a $\SO(6)$ NGB-dependent rotation on $\Psi_{R}$, it becomes explicit that either $M_{14}$, $M_{5}$, or $M_1$, are also needed. 

After mixing (at zeroth order in $h$ and $\eta$) the $q_L$ and $t_R$ states with the resonances in $\Psi$ with the proper gauge quantum numbers, the top Yukawa coupling in \Eq{eq:topYuk20} is generated,
\beq
y_t = \frac{ \sqrt{12/5} \, \lambda_L \lambda_R \, f M_{14} M_{5}}{\sqrt{M_{14}^2+(\hat \gamma \lambda_L f)^2} \sqrt{M_5^2 + (\lambda_L f)^2} \sqrt{M_1^2 + (\lambda_R f)^2}} \ .
\label{eq:Yukcoupling20}
\eeq
The presence of two different couplings for $q_L$, one of them proportional to $\hat \gamma$, introduces two mixing angles for $t_L$. 
Recalling that $\lambda_R$ does not introduce any explicit breaking of $\SO(6)$, already from \Eq{eq:Yukcoupling20} one can see that the regime $M_1 \ll \lambda_R f$ will be preferred. 
This allows $\lambda_L$, which controls the size of the top sector contribution to the pNGB potential, to be the smallest possible compatible with the large top Yukawa, $\lambda_L \simeq y_t \sqrt{5/12}$.
From now on we will take $M_1 = 0$. 
The mass of the singlet top partner is then $m_{\Psis} = \lambda_R f$, and given that it is not associated to any breaking, it drops out completely from the potential. 

Indeed, the computation of the Coleman-Weinberg potential gives rise to the following coefficients $c^{(1)}_L$ and $c^{(2)}_L$ in \Eqs{eq:VL2a} and (\ref{eq:VL2b}):
\beq
\label{eq:c2ab}
c^{(1)}_L = \frac{3}{4 \pi^2} \left( \tilde M_5^2 - \tilde M_{14}^2 \right)  \ , \quad 
c^{(2)}_L = \frac{3}{20 \pi^2} \left( 2 \tilde M_{14}^2 - 5 \tilde M_5^2 \right) \ ,
\eeq
where we have defined $\tilde M_{5, 14}^2 \equiv M_{5, 14}^2 \log (\Lambda^2 / M_{5,14}^2 )$. 
Taking into account \Eq{eq:Yukcoupling20}, and keeping the leading order terms $\order{\lambda_L^2}$ only, this model predicts, 
\bea
\label{mH220p}
(\Delta \mu^2)_L
\!\!\! &\simeq& \!\!\!
\frac{3 y_t^2}{96 \pi^2} \left[ 5 \tilde M_{5}^2 (7 - \hat \gamma^2) - \tilde M_{14}^2 (23 - 5 \hat \gamma^2) \right] \ , \\
\label{lambda20}
\lambda
\!\!\! &\simeq& \!\!\!
\frac{3 y_t^2}{12 \pi^2} \frac{5 \tilde M_{5}^2 - 2 \tilde M_{14}^2}{f^2} \ ,
\eea
for the Higgs mass term and quartic. 
Both arise at leading order, and they are sensitive to a second level of resonances through $\Lambda$, which have been reabsorbed in the effective masses $\tilde M_{5, 14}$. 
The mass of the singlet is given by
\beq
(\Delta m_\eta^2)_L \simeq (\hat \gamma^2-1) \frac{15 y_t^2}{24 \pi^2} \left( \tilde M_{5}^2 - \tilde M_{14}^2 \right) \ .
\label{etamass20p}
\eeq
As we advanced, in the limit $\hat \gamma \to 1$ the singlet does not receive a potential from the top sector. 
Notice also that regardless of $\hat \gamma$, $\eta$ becomes massless in limit $M_{5} \to M_{14}$ ($c^{(1)}_L \to 0$ in this limit), due to an enhanced global symmetry of the top Lagrangian \Eq{eq:Ltop20p}.

If the top sector gives the only relevant contribution to the Higgs quartic, then \Eq{lambda20} implies a particular relation between $\tilde M_{5}$ and $\tilde M_{14}$, which is satisfied without fine tuning as long as $|\tilde M_{5}/f| \lesssim 1.5$ and $|\tilde M_{14}/f| \lesssim 2.5$. 
Therefore a light Higgs requires weakly coupled top partners. 
The relation between $\tilde M_{5}$ and $\tilde M_{14}$ enforced by $\lambda$ fixes also the level of tuning in the Higgs mass term, as well as the mass of the singlet, as a function of a single top partner's mass parameter,
\beq
\mu^2 \simeq (3 - \hat \gamma^2) \frac{3 y_t^2 \tilde M_{14}^2}{32 \pi^2} + \frac{\lambda}{8} (7 - \hat \gamma^2) f^2 + (\Delta \mu^2)_{g} \ , \quad m_\eta^2 \simeq (\hat \gamma^2 -1) \frac{4 \pi^2 \lambda f^2 - 3 y_t^2 \tilde M_{14}^2}{8 \pi^2} \ .
\label{eq:masses201}
\eeq
This is neglecting other contributions, such as \Eq{eq:technimass}, as well $\order{y_t^4}$ terms. 
It is important for the phenomenology of the top partners (see section~\ref{pheno}) to discuss the role of $\hat \gamma$. 
For $\hat \gamma < 1$, there is a lower bound on $\tilde M_{14}^2$ such that $m_\eta^2$ is positive (which corresponds to $\tilde M_{14}^2 \geqslant \tilde M_{5}^2$). 
For $\hat \gamma > 1$, this becomes an upper bound. 
Naively $\hat \gamma > 1$ is preferred, given that it diminishes the contributions to $\mu^2$, for fixed $\tilde M_{14}^2$, and thus reduces the level of fine tuning. 
However, for $\hat \gamma \neq 0$ there is a theoretical lower bound on $M_{14}$ from the requirement to reproduce the top Yukawa, $(M_{14})_{min} = \hat \gamma \sqrt{5/12} \, y_t f$ (as well as $(M_{5})_{min} = \sqrt{5/12} \, y_t f$), which forces large values of $M_{14}$, increasing the tuning for both the Higgs mass term and quartic.
This lower bound on $M_{14}$ has to be compared with the upper bound on $\tilde M_{14}^2$ enforcing $m_\eta^2 > 0$. 
However, this comparison relies on the scale $\Lambda$, and therefore we cannot directly establish the consistency of these two limits without an explicit model that eliminates this lack of calculability (such as a 3-site model). 
We can nevertheless say that $\hat \gamma \gtrsim 1$, but not excessively large, is preferred in this model. 
We can also give a simple estimate of the top sector contribution to the mass of $\eta$ by setting $\tilde M_{14} = 0$ in \Eq{eq:masses201},
\beq
\Delta m_\eta^2 \sim (\hat \gamma^2 -1) \frac{m_h^2 f^2}{4 v^2} \simeq (230 \GeV)^2 \left( \frac{|\hat \gamma^2-1|}{2} \right) \left( \frac{15\%}{v^2/f^2} \right) \ .
\eeq

\begin{figure}[!t]
\begin{center}
\includegraphics[width=3in]{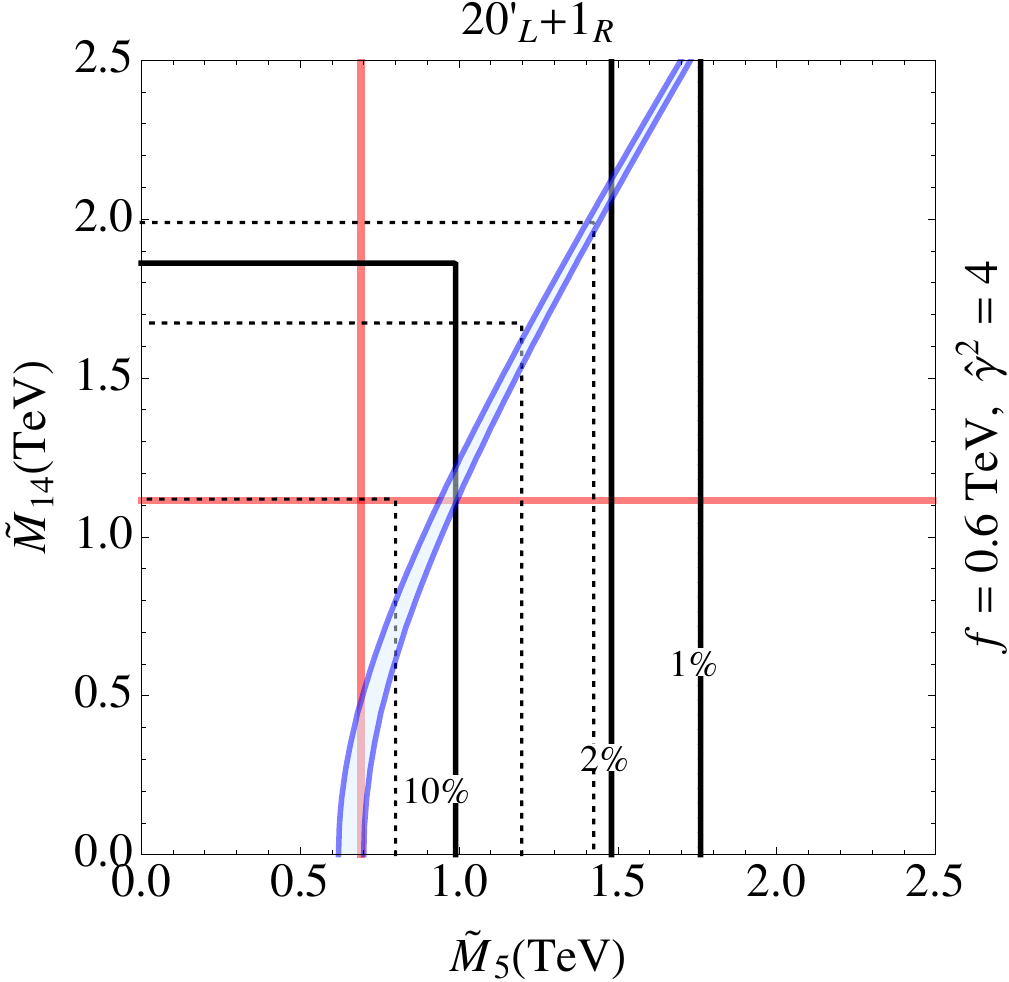}
\caption{Contour lines of tuning as defined in the text (solid black) and regions with Higgs quartic $0.11 \leqslant \lambda \leqslant 0.14$ (blue), in the plane of the top partners mass parameters $\tilde M_5$ and $\tilde M_{14}$, with $f= 0.6 \TeV$ and $\hat \gamma^2 = 4$.
Contour lines of tuning for $\hat \gamma = 0$ (dotted black) are also shown for comparison. 
The red lines delimitate the region (upper-right) where the top Yukawa can be reproduced,
and we have simply taken $\Lambda = 2.5 \TeV$.
For $\hat \gamma = 0$ the lower bound on $M_{14}$ goes to zero.}
\label{fig:tuningmass20p}
\end{center}
\end{figure}

To illustrate the points discussed above, we show in Figure \ref{fig:tuningmass20p} contour lines for the tuning in this model, as well as the region where the Higgs quartic is reproduced, in the plane ($\tilde M_{5}, \tilde M_{14})$, and for a representative set of parameters. 
We defined the tuning as $\max[ (\partial \log \mu^2/\partial \log \tilde M_{i}^2) (\partial \log \lambda/\partial \log \tilde M_{i}^2)]$, for $\tilde M_{i} = \tilde M_{5}, \tilde M_{14}$, thus treating as separate contributions to the potential the terms proportional to $\tilde M_{5}^2$ and those proportional to $\tilde M_{14}^2$. 

\section{Non-Minimal Top Partner Phenomenology} \label{pheno}

We have explicitly shown in the previous section that in the NMCHM the masses of the top partners control the size of the Higgs potential. 
The mass of the extra singlet $\eta$ also gets contributions from the top partners, such that $\Delta m_\eta^2 \sim (\gamma^2 -1) m_\Psi^2$. 
The extra symmetry breaking coupling associated to $\gamma$ (or $\hat \gamma$ depending on the embedding of the top) does not modify significantly the predictions for the Higgs potential, and in particular $\gamma = \order{1}$ does not give rise to a larger level of tuning. 
Interestingly, we will show in this section that the decay channel $\Psis \to \eta t$ becomes important when $\gamma \neq 0$, where $\Psis$ is the top partner in the singlet representation of $\SO(5)$.\footnote{Other than the extra decay to $\eta t$, $\Psis$ has the same main characteristics as the top partner singlet of $\SO(4)$ in the $\SO(5)/\SO(4)$ model, denoted by $\tilde T$ in \cite{DeSimone:2012fs}.} 
We will also show that one of the top partners belonging to the $\mathbf{5}$ multiplet of $\SO(5)$ decays exclusively to $\eta t$.\footnote{This is also the case for some of the top partners in the $\mathbf{14}$, although we will not discuss them here.} 
In order to arrive at such results, we will derive, for a single $\SO(5)$ multiplet of top partners at a time, its interactions with the NGBs and the SM fields.
We will make use of effective Lagrangians that implicitly assume that other composite resonances, in particular other multiplets of top partners, are heavy and lie at or beyond the cutoff. 
Such type of Lagrangians must be invariant under local $\SO(5)$ transformations, thus reproducing the non-linearly realized $\SO(6)$ symmetry of the strong sector. 
Its building blocks are 
\emph{i)} the top partners, belonging to a given $\SO(5)$ multiplet (and with a definite $X$ charge, for the cases considered here equal to 2/3), 
\emph{ii)} the derivatives of the NGBs, introduced through $d_\mu^{\ahat} = -i \Tr[T^{\ahat} U^\dagger D_\mu U]$ and the $\SO(5)$ gauge connection $e_\mu^a =  -i \Tr[T^a U^\dagger D_\mu U]$, with $T^{\ahat}$ and $T^a$ the broken and unbroken generators of $\SO(6)$ respectively (see Appendix~\ref{ccwz} for details), and 
\emph{iii)} the SM states, specifically the $\SU(3)_C \times \SU(2)_L \times \UU(1)_Y$ gauge fields and the top quark. 
Regarding the latter, recall that we specified its embedding in $\SO(6)$ representations, \Eqs{eq:embed6L} and (\ref{eq:embed6R}) or \Eq{eq:embed20L}. 
In order to include them in our effective Lagrangian, we will use the NGB matrix $U$ to form the dressed fields $U^{i}_{\ I} (Q^{\mathbf{6}}_L)^I$ and $U^{6}_{\ I} (Q^{\mathbf{6}}_L)^I$ transforming as a $\mathbf{5}$ and a $\mathbf{1}$ under $\SO(5)$ respectively, and likewise for $T^{\mathbf{6}}_R$ and $Q^{\mathbf{20'}}_L$.
This approach, also followed in \cite{DeSimone:2012fs}, is very efficient in systematically identifying the leading interactions of the top partners, in an expansion in derivatives and symmetry breaking couplings. 


\subsection{Singlet phenomenology} \label{1pheno}

Let us focus first on the phenomenology of the top partner singlet of $\SO(5)$, $\Psis$, for the case where both the $q_L$ and $t_R$ are embedded in the $\mathbf{6}_{\mathbf{2/3}}$ of $\SO(6) \times \UU(1)_X$. 
The effective Lagrangian, at leading order in derivatives and elementary-composite couplings $y_{L,R}$, is
\footnote{We are neglecting terms at the same order in $y_L$ and $y_R$ but with one extra derivative, given that these are effectively suppressed by $g_\Psi/g_\rho$, where $g_\rho$ is a strong coupling associated to heavy composite states, in the sense $g_\rho \gg g_\Psi \equiv M_\Psi/f $.}
\bea
\Lag^{\mathbf{6}_L+\mathbf{6}_R}_{\Psis}
\!\!\! &=& \!\!\! i \, \bar{q}_L \Dslash q_L + i \, \bar{t}_R \Dslash t_R + i \, \bar{\Psi}^{\mathbf{1}} \Dslash \Psis - M_{\Psi} \bar{\Psi}^{\mathbf{1}} \Psis \nn \\
\!\!\! &-& \!\!\! y_L f (\bar Q^{\mathbf{6}}_L)_I U^I_{\ 6} \Psis_R - y_R f \bar \Psi^{\mathbf{1}}_L U^{6}_{\ I} (T^{\mathbf{6}}_R)^I + h.c. \ .
\label{eq:Lsinglet66}
\eea
The parameters of this Lagrangian can be taken to be real without any loss of generality and consistently with $\cp$ conservation. 
The covariant derivatives acting on $q_L$ and $t_R$ encode the usual SM gauge interactions, and given that $\Psis$ has hypercharge $Y=X=2/3$ and it is a color fundamental (the same gauge quantum numbers as $t_R$), its covariant derivative contains also the corresponding gauge connections. 
Importantly, only the last two terms in \Eq{eq:Lsinglet66} depend on the NGBs, and in a non-derivative way. 
Both of them induce a mixing between the $\Psis$ and the SM top, but only the term proportional to $y_R$ does it at leading order in $v/f$ (recall $\vev{h} = v$).
Then the masses of the top and the top partner (we use the same notation before and after rotation to the mass basis) read
\beq
m_t \simeq \frac{y_L y_R}{\sqrt{g_\Psi^2 + y_R^2}} \frac{v}{\sqrt{2}} \ , \quad 
m_{\Psis} \simeq f \sqrt{g_\Psi^2 + y_R^2}
\label{eq:masses1}
\eeq
where we defined $g_\Psi \equiv M_\Psi/f$, and neglected subleading $\order{y_{L,R}^2 v^2 / g_\Psi^2 f^2}$ terms.

\begin{figure}[!t]
\begin{center}
\includegraphics[width=3.1in]{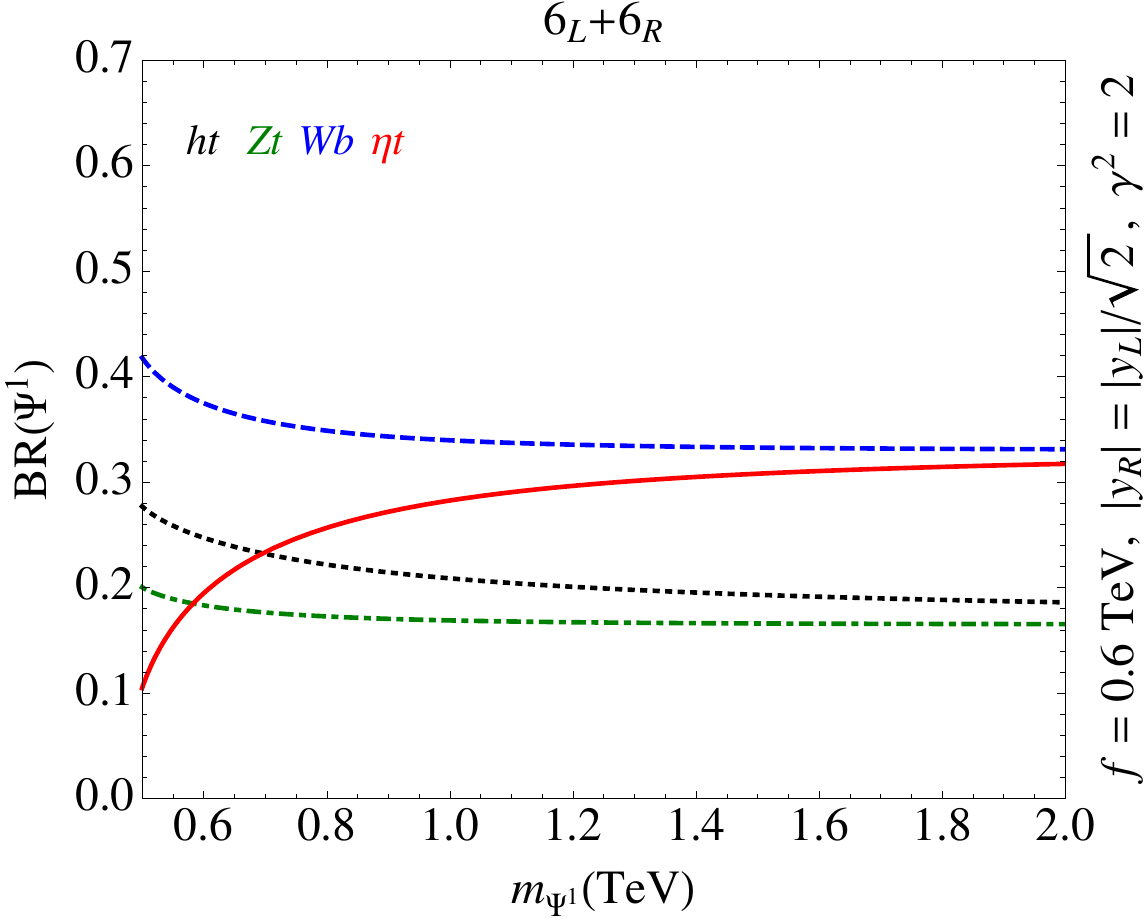}
\hspace{7pt}
\includegraphics[width=3.1in]{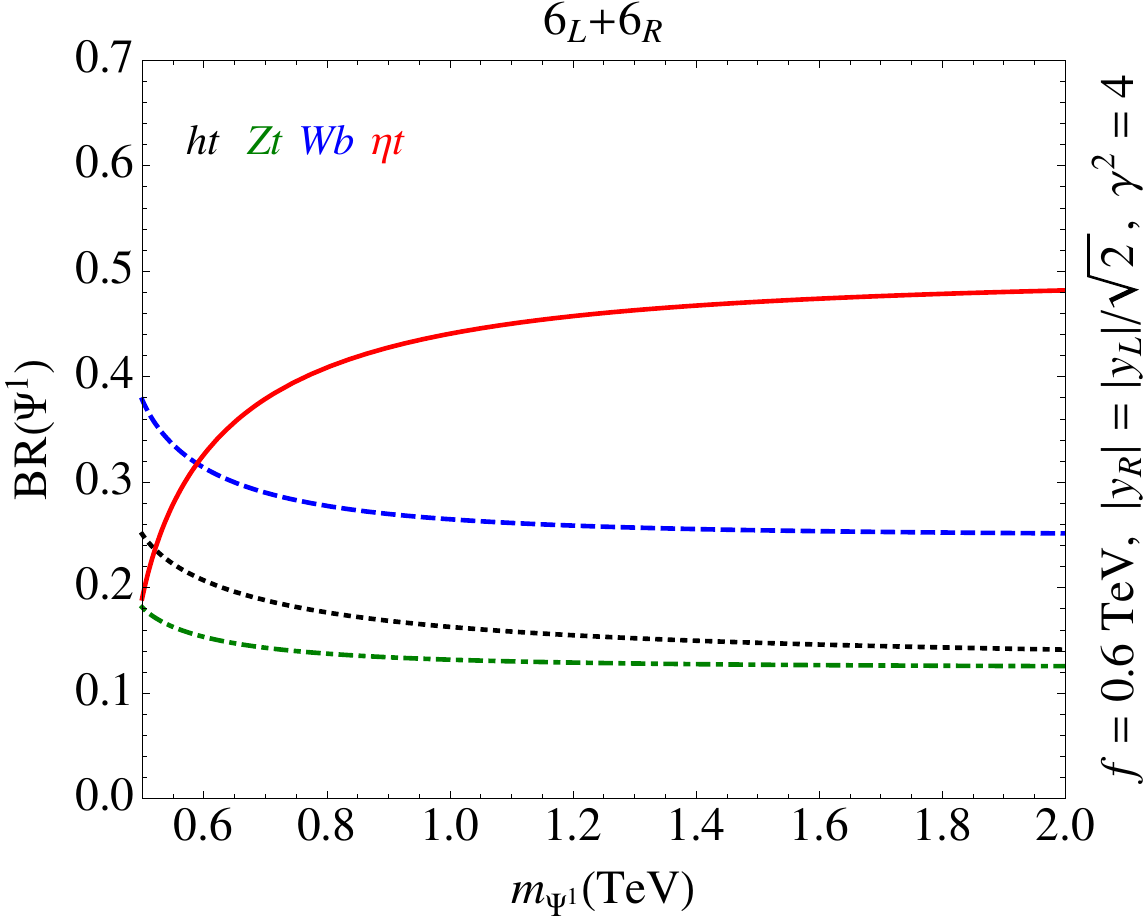}
\caption{Branching ratios of $\Psis$ to $ht$ (dotted black), $Zt$ (dot-dashed green), $Wb$ (dashed blue), and $\eta t$ (solid red), in the $\mathbf{6}_L + \mathbf{6}_R$ model. 
The left panel correspond to $\gamma^2 = 2$ and the right one to $\gamma^2 = 4$. 
The singlet mass has been fixed in both cases to $m_\eta = 300 \GeV$, while $y_R = y_L/\sqrt{2}$.}
\label{fig:brsinglet66}
\end{center}
\end{figure}

The most relevant interactions for what regards the decays of $\Psis$ (and its single production) come from the trilinear couplings between a top partner, a third generation SM quark, and a single NGB, either the physical Higgs scalar $h$,\footnote{With a slight abuse of notation, we will denote with $h$ also the scalar fluctuation around the electroweak VEV, that is $h \to v + h$.} the pseudo-scalar $\eta$, or the longitudinal components of the $W^\pm$ and the $Z$. 
The latter are, by the equivalence theorem, well approximated by the Goldstone degrees of freedom in the Higgs field, $\phi^\pm$ and $\phi^0$, respectively. 
We will consider only the leading couplings arising at zeroth order in $v/f$. 
We will therefore neglect the interactions with the transverse components of the $W^\pm$ and the $Z$, given that these are diagonal in flavor space, and only after EWSB they give rise to a coupling of a SM quark and a top partner. 
Besides, the electroweak gauge couplings $g$ or $g'$ are smaller than the Yukawa-type couplings, proportional to $y_L, y_R \gtrsim y_t$. 
Under these approximations, the linear couplings of the top partner $\Psis$ are
\beq
\frac{g_\Psi y_L}{\sqrt{g_\Psi^2 + y_R^2}} \left[ \frac{1}{\sqrt{2}} (h - i \phi_0) \bar t_L \Psis_R - \phi^- \bar b_L \Psis_R \right]
- i \, \frac{g_\Psi y_R \gamma}{\sqrt{g_\Psi^2 + y_R^2}} \, \eta \bar \Psi^{\mathbf{1}}_L t_R + h.c. \ .
\eeq
These then imply the following approximate relations between the branching ratios of $\Psis$:
\beq
\frac{\BR(\Psis \to \eta t)}{\BR(\Psis \to h t)} \simeq
\frac{\BR(\Psis \to \eta t)}{\BR(\Psis \to Z t)} \simeq
2 \frac{\BR(\Psis \to \eta t)}{\BR(\Psis \to W^+ b)} \simeq
\frac{2 y_R^2 \gamma^2}{y_L^2} \ ,
\label{eq:brsinglet66}
\eeq
where we have neglected kinematical factors, assuming in particular that $m_{\Psis} \gg m_\eta + m_t$.\footnote{We have also neglected in \Eq{eq:brsinglet66} the rescaling of all the couplings of $h$ from the correction to its kinetic term in \Eq{eq:kinetic}.} 
Therefore, the decay channel $\Psis \to \eta t$ becomes important with $\gamma^2$. 
As an example, we can match the parameters in the effective Lagrangian \Eq{eq:Lsinglet66} to the model presented in section \ref{model1}, after integrating out the 5-plet $\Psif$. 
One then obtains
\beq
g_\Psi = \lambda_\Psi - y_\Psi = M_1/f \, , \quad y_R = \lambda_R \, , \quad y_L = \lambda_L \ ,
\eeq
and consequently, 
\beq
\BR(\Psis \to \eta t) \simeq 1 - \frac{1}{1+(\gamma \zeta)^2/2} \ ,
\eeq
where $\zeta = \lambda_R/\lambda_L$. 
Recalling that from fine-tuning considerations the regime $\zeta \simeq 1/\sqrt{2}$ was preferred in that simple model, then $\BR(\Psis \to \eta t) \simeq 33 \%$ for $\gamma^2 = 2$, a branching ratio as large as that into $W^+ b$ (which is the dominant channel in the $\SO(5)/\SO(4)$ model). 
In Figure \ref{fig:brsinglet66} we show the branching ratios of $\Psis$ as a function of its mass, for $y_R = y_L/\sqrt{2}$ and $\gamma^2 = 2$ or $4$. 
We have fixed in both cases $m_\eta = 300 \GeV$, to illustrate the fact that if there are no kinematical suppressions the decay $\Psis \to \eta t$ can dominate. 
Notice however that if $\gamma$ becomes very large, such a decay is only kinematically allowed for a heavy $\Psis$, given that the mass of $\eta$ grows with $\gamma^2$ (and we do not wish to tune down $m_\eta$). 
Furthermore, recall that for increasing $\gamma$ the theoretical lower bound on the mass of the 5-plet of top partners also grows. 
In summary, the non-standard $\eta t$ decay can dominate, but not at the level of making the other decays negligible. 

The phenomenology of $\Psis$ for the case where the $q_L$ and $t_R$ are embedded respectively in the $\mathbf{20'}_{\mathbf{2/3}}$ and $\mathbf{1}_{\mathbf{2/3}}$ can be described a similar way.  
The leading terms in the corresponding effective Lagrangian are
\bea
\Lag^{\mathbf{20'}_L+\mathbf{1}_R}_{\Psis}
\!\!\! &=& \!\!\! i \, \bar{q}_L \Dslash q_L + i \, \bar{t}_R \Dslash t_R + i \, \bar{\Psi}^{\mathbf{1}} \Dslash \Psis - M_{\Psi} \bar{\Psi}^{\mathbf{1}} \Psis\nn \\
\!\!\! &+& \!\!\! y_L f U^I_{\ 6}(\bar Q^{\mathbf{20'}}_L)_{I J} U^J_{\ 6} \Psis_R + y_L c_t f U^I_{\ 6}(\bar Q^{\mathbf{20'}}_L)_{I J} U^J_{\ 6} t_R + h.c. \ .
\label{eq:Lsinglet201}
\eea
The Yukawa coupling of $t_R$ does not need to be suppressed with respect to that of $\Psis_R$, therefore $c_t = \order{1}$. 
Only the terms in the second line depend on the NGBs: at leading order in $v/f$, the last gives rise to the top mass, $m_t \simeq y_L c_t v$,  while the first gives rise to the leading non-diagonal interactions of $\Psis$ with $q_L$. 
The mass of $\Psis$ at this order is simply $M_\Psi$. 
Here we will once again neglect the couplings to transverse gauge bosons, subleading in the $v/f$ expansion and also because $g, g' < y_L \simeq y_t$. 
Then the relevant interactions are
\beq
-y_L \left[ \frac{1}{\sqrt{2}} (h - i \phi_0) \bar t_L \Psis_R - \phi^- \bar b_L \Psis_R + i \hat \gamma \frac{v}{f} \eta \bar t_L \Psis_R \right] + h.c. \ .
\eeq
Notice that because of $\SU(2)_L$ quantum numbers, the leading coupling of $\Psis$ to $\eta$ arises at order $v/f$, but it is enhanced by $\hat \gamma$. 
The branching ratios of $\Psis$ are then:
\beq
\frac{\BR(\Psis \to \eta t)}{\BR(\Psis \to h t)} \simeq
\frac{\BR(\Psis \to \eta t)}{\BR(\Psis \to Z t)} \simeq
2 \frac{\BR(\Psis \to \eta t)}{\BR(\Psis \to W^+ b)} \simeq
\hat \gamma^2 \frac{v^2}{f^2} \ .
\label{eq:brsinglet201}
\eeq
Therefore a large non-standard branching ratio requires significantly large values of $\hat \gamma$, to overcome the $v^2/f^2$ suppression. 
For instance, given $\hat \gamma^2 = 4$ and $f = 600 \GeV$, one finds $\BR(\Psis \to \eta t) \simeq 20 \%$.
We can understand the feasibility of this regime by matching the effective Lagrangian \Eq{eq:Lsinglet201} to the model presented in section \ref{model2}, assuming that the top partners in the 5-plet and the 14-plet are heavy. 
In that case we find
\beq
M_\Psi = \sqrt{M_1^2 + (\lambda_R f)^2} \, , \quad y_L = - \sqrt{\frac{6}{5}} \frac{\lambda_L M_1}{\sqrt{M_1^2 + (\lambda_R f)^2}} \, , \quad c_t = - \frac{\lambda_R f}{M_1} \ .
\eeq
Recall in particular that the regime $M_1/f \ll \lambda_R$ was preferred for fine-tuning considerations, in which case $M_1$ was playing little role in the Higgs potential. 
Besides, even though large values of $\hat \gamma$ were preferable, the theoretical lower limit on $M_{14}$ scaled also with $\hat \gamma$, possibly becoming the leading source of tuning for $\hat \gamma \gg 1$. 
Therefore, we can conclude that a non-standard branching ratio of $\Psis$ to $\eta t$ can become comparable to those into the standard channels, but not dominant.

\subsection{5-plet phenomenology} \label{5pheno}

We discuss next the phenomenology of the top partner 5-plet of $\SO(5)$, $\Psif$. 
For the case where both the $q_L$ and $t_R$ are embedded in the $\mathbf{6}_\mathbf{2/3}$ of $\SO(6) \times \UU(1)_X$, the leading effective Lagrangian reads
\bea
\Lag_{\Psif}^{\mathbf{6}_L+\mathbf{6}_R} 
\!\!\! &=& \!\!\! i \, \bar{q}_L \Dslash q_L + i \, \bar{t}_R \Dslash t_R + i \, \bar \Psi^{\mathbf{5}} (\Dslash + i e_\mu \gamma^\mu) \Psif - M_\Psi \bar \Psi^{\mathbf{5}} \Psif \nn \\
\!\!\! &-& \!\!\! y_L f (\bar Q_L)_I U^I_{\ i} (\Psif_R)^i - y_R f (\bar \Psi^{\mathbf{5}}_L)_i U^{i}_{\ I} (T_R)^I + h.c. \ .
\label{eq:L5plet66}
\eea
Its parameters can be taken to be real, consistently with $\cp$ conservation and without loss of generality. 
The covariant derivative acting on $\Psif$ contains the gauge connection associated with the $X=2/3$ charge of the whole multiplet, as well as the color gauge connection, such that $D \Psif = (\partial - i g' (2/3) B - i g_s G) \Psif$.
The kinetic term for $\Psif$ contains also the object $e_\mu$, required by the local $\SO(5)$ symmetry. 
Such a term encondes the proper electroweak gauge interactions of the components in the 5-plet: $(T, B) = \mathbf{2}_{\mathbf{1/6}}$, $(X_{5/3}, X_{2/3}) = \mathbf{2}_{\mathbf{7/6}}$ and $T' = \mathbf{1}_{\mathbf{2/3}}$ under $\SU(2)_L \times \UU(1)_Y$, where we recall that $Y = T_R^3 + X$. 
The masses of these top partners, after accounting for the mixing with $q_L$ and $t_R$, are
\beq
 m_{X_{5/3}} = m_{X_{2/3}} = M_\Psi \equiv g_\Psi f
\, , \quad
m_B \simeq m_T \simeq f \sqrt{g_\Psi^2 + y_L^2}
\, , \quad
 m_{T'} \simeq \sqrt{g_\Psi^2 + (\gamma y_R)^2} \ ,
\label{eq:masses566}
\eeq
while the top mass is
\beq
m_t \simeq \frac{y_L y_R g_\Psi}{\sqrt{g_\Psi^2 + y_L^2} \sqrt{g_\Psi^2 + (\gamma y_R)^2}} \frac{v}{\sqrt{2}} \ .
\eeq
In these expressions we neglected EWSB corrections, effectively suppressed by $y_{L,R}^2 v^2 /g_\Psi^2 f^2 $.

The decays of the top partners in the 5-plet are mostly determined by the trilinear interactions from the second line in \Eq{eq:L5plet66}, which involve a NGB and a third generation SM quark. 
The interactions with the transverse gauge bosons, from the first line in \Eq{eq:L5plet66}, are effectively suppressed by $v/f$ and $g/y_{L,R}$, and we will neglect them in what follows (this is in analogy with the interactions of the singlet $\Psis$ in section \ref{1pheno}). 
Under these approximations, the relevant couplings of the top partners in $\Psif$ are
\beq
y_R c_{\Psi_R} \left[
\frac{c_{\Psi_L}}{\sqrt{2}} (h - i \phi_0) \bar T_L -
\frac{1}{\sqrt{2}} (h + i \phi_0) \bar X_{2/3 \, L} -
c_{\Psi_L} \phi^- \bar B_L -
\phi^+ \bar X_{5/3 \, L} +
i \eta \bar T'_L \right] t_R + h.c. \ ,
\label{eq:c566}
\eeq
where $c_{\Psi_R} \equiv g_\Psi/\sqrt{g_\Psi^2 + \gamma^2 y_R^2}$ and $c_{\Psi_L} \equiv g_\Psi/\sqrt{g_\Psi^2 +y_L^2}$. 
Notice first that all these interactions are proportional to $y_R$. 
The Yukawa-type invariant proportional to $y_L$ in \Eq{eq:L5plet66} does not give rise to couplings of the $\SU(2)_L$ doublets in $\Psif$ with $q_L$ unless the electroweak symmetry is broken (thus they arise at order $v/f$), and likewise for the singlet $T'$. 
Furthermore, the trilinear couplings $\phi_0 \bar B b$, $\phi^+ \bar X_{2/3} b$ are absent, in analogy with the same couplings in the $\SO(5)/\SO(4)$ model (see \cite{DeSimone:2012fs} for details).
Finally, the interactions $\phi_+ \bar T' b$, $(h-i\phi_0) \bar T' t$, $\eta \bar T t$, and $\eta \bar X_{2/3} t$ are also vanishing. 
This is due to a parity symmetry $P_\eta$: $H \to H, \, \eta \to - \eta$ and  $T, B, X_{2/3}, X_{5/3} \to T, B, X_{2/3}, X_{5/3}, \, T' \to - T'$, which is preserved by the leading interactions in the effective Lagrangian \Eq{eq:L5plet66}. 
Summarizing, the interactions in \Eq{eq:c566} imply the following branching ratios for the $\Psif$ components:
\bea
& \BR(T, X_{2/3} \to h t) \simeq \BR(T, X_{2/3}\to Z t) \simeq 50 \% \ , & \nn \\
& \BR(B \to W^- t) \simeq \BR(X_{5/3} \to W^+ t) \simeq \BR(T' \to \eta t) \simeq 100 \% \ , &  
\label{eq:br5plet66}
\eea
neglecting kinematical factors. 
Interestingly, the extra top partner $T'$, associated to the larger $\SO(6)$ symmetry of the NMHCM (compared with the $\SO(5)$ of the minimal model), decays exclusively to the extra NGB $\eta$. 
The signatures at colliders from the $T'$ could provide important indications towards such an extended symmetry structure. 

Let us comment on one more possibility regarding the decays of the 5-plet, still for the $q_L$ and $t_R$ being embedded in the $\mathbf{6}$.
One of the conclusions extracted from the model presented in section \ref{model1} was that for $\gamma \neq 0$ $\Psif$ must be generically heavier than the singlet $\Psis$. 
That being the case, we may wonder how the decays of the 5-plet would change by including $\Psis$ in the effective Lagrangian. 
Given that both of them are composite states, their interactions would be stronger than those with the elementary $q_L$ or $t_R$. 
Indeed, at leading order in derivatives we can add to the effective Lagrangian the term
\beq
i \, c_L \, (\bar \Psi^{\mathbf{5}}_L)_i \, d^i_\mu \gamma^\mu \Psis_L + h.c. + L \leftrightarrow R \ ,
\label{eq:L5singlet}
\eeq
with $c_L, c_R = \order{1}$ and real. 
Within the assumption that the 5-plet is heavier than the singlet, we will keep only the leading order interactions in $g_{\Psif} = M_{\Psif}/f$. 
This implies in particular that we will neglect the mixings of $q_L$ and $t_R$ with $\Psif$. 
We must also keep in mind that the interactions in \Eq{eq:L5singlet} involve derivatives of the NGBs, which after integrating by parts give rise to couplings proportional to the masses of the top partners. 
Then the relevant trilinear interactions, coming from \Eq{eq:L5singlet}, are
\beq
c_R \frac{g_{\Psif} y_R}{\sqrt{g_{\Psis}^2 + y_R^2}} \bar T_L (h - i \phi^0) t_R + h.c. \ , 
\eeq
\beq
\left( c_L \sqrt{g_{\Psis}^2 + y_R^2}
- c_R \frac{g_{\Psif} g_{\Psis}}{\sqrt{g_{\Psis}^2 + y_R^2}}  \right) \bar T_L (h - i \phi^0) \Psis_R +
\left(c_R g_{\Psis} - c_L g_{\Psif} \right) \bar T_R (h - i \phi^0) \Psis_L
 + h.c. \ ,
\eeq
for the $T$ top partner, and where $y_R$ is the Yukawa coupling of $t_R$ with $\Psis$ in \Eq{eq:Lsinglet66}.
The couplings of the rest of top partners in $\Psif$ can be easily obtained by making the substitutions $T \to X_{2/3}$ along with $(h-i\phi^0) \to - (h+i\phi^0)$, $T \to T'$ with $(h -i\phi^0) \to \sqrt{2} i \eta$, $T \to X_{5/3}$ with $(h-i\phi^0) \to \sqrt{2} \phi^+$, and $T \to B$ with $(h-i\phi^0) \to -\sqrt{2} \phi^-$. 
Even thought the decay to $t_R$ is still relevant, as long as $g_{\Psis} > y_R$ the decays of the 5-plet to $\Psis$ easily dominate: 
\beq
\frac{\BR(\Psif \to \Psis \, \Pi)}{\BR(\Psif \to t \, \Pi)} \simeq \frac{c_L^2 (g_{\Psis}^2 + y_R^2) + c_R^2 g_{\Psis}^2}{c_R^2 y_R^2} \ , 
\eeq
under the assumption that the decay is kinematically allowed. 
Finally, notice that once again $T'$ decays exclusively to $\eta$. 

The phenomenology of $\Psif$ when the $q_L$ and $t_R$ are embedded respectively in the $\mathbf{20'}_{\mathbf{2/3}}$ and $\mathbf{1}_{\mathbf{2/3}}$ is described by the effective Lagrangian
\bea
\label{eq:L5plet201}
\Lag^{\mathbf{20'}_L+\mathbf{1}_R}_{\Psif}
\!\!\! &=& \!\!\! i \, \bar{q}_L \Dslash q_L + i \, \bar{t}_R \Dslash t_R + i \, \bar \Psi^{\mathbf{5}} (\Dslash + i e_\mu \gamma^\mu) \Psif - M_\Psi \bar \Psi^{\mathbf{5}} \Psif \\
\!\!\! &+& \!\!\! i \, c_R \, (\bar \Psi^{\mathbf{5}}_R)_i \, d^i_\mu \gamma^\mu t_R + 2 \, y_L f U^I_{\ 6}(\bar Q^{\mathbf{20'}}_L)_{I J} U^J_{\ i} (\Psif_R)^i + y_L c_t f U^I_{\ 6}(\bar Q^{\mathbf{20'}}_L)_{I J} U^J_{\ 6} t_R + h.c. \ , \nn
\eea
which contains a term like \Eq{eq:L5singlet} but with $t_R$ instead of $\Psis_R$. 
Given that the $R$-handed top interacts like a singlet of $\SO(6)$, it may couple strongly to $\Psif$ without inducing a large Higgs potential. 
Therefore, the trilinear interactions from the first term in the second line of \Eq{eq:L5plet201}, with $c_R = \order{1}$, will generically dominate over the interactions from the second term. 
Under this assumption, we find that the relevant couplings of $\Psif$ are
\bea
&& c_R \sqrt{g_{\Psi}^2 + y_L^2} \left[
- (h - i \phi_0) \bar T_L +
\sqrt{2} \phi^- \bar B_L 
\right] t_R \nn \\
&+& c_R \, g_\Psi \left[
(h + i \phi_0) \bar X_{2/3 \, L} + 
\sqrt{2} \phi^+ \bar X_{5/3 \, L} -
i \sqrt{2} \eta \bar T'_L
\right] t_R + h.c. \ ,
\label{eq:c520p}
\eea
where $g_\Psi \equiv M_\Psi/f$, and the masses of the top partners are $m_{X_{5/3}} \simeq m_{X_{2/3}} \simeq m_{T'} \simeq g_\Psi f$, and $m_B \simeq m_T \simeq f \sqrt{g_\Psi^2 + y_L^2}$, while the mass of the top is $m_t \simeq y_L c_t v$. 
Therefore we find the branching ratios:
\bea
& \BR(T, X_{2/3} \to h t) \simeq \BR(T, X_{2/3}\to Z t) \simeq 50 \% \ , & \nn \\
& \BR(B \to W^- t) \simeq \BR(X_{5/3} \to W^+ t) \simeq \BR(T' \to \eta t) \simeq 100 \% \ . &  
\label{eq:br5plet20p}
\eea
These are the same branching ratios as in the case with $q_L$ and $t_R$ embedded in the $\mathbf{6}$ of $\SO(6)$, \Eq{eq:br5plet66}, even though in this case they arise from Lagrangian terms with derivatives acting on the NGBs. 
This means that while in the previous case the presence of a light singlet $\Psis$ to which $\Psif$ could decay to would easily dominate the branching ratios, in the present case $t_R$ could be as strongly coupled as a hypothetically light $\Psis$, and the decay channels $\Psif \to t \, \Pi$ and $\Psif \to \Psis \, \Pi$ would generically be comparable.


\section{Goldstone-Singlet Phenomenology} \label{phenosinglet}

The non-standard phenomenology of the top partners in the NMCHM relies on the extra Goldstone mode $\eta$. 
Specifically, the final state particles in the production and subsequent decay of $\Psis$ and $T'$ will be ultimately determined by the decay products of $\eta$. 
Besides, understanding the phenomenology at colliders of the pseudo-scalar singlet is important per se. 
This is the aim of this section. 

Before we do so, let us briefly comment on the couplings of the Higgs. 
These are modified with respect to the SM ones mainly because of the non-linearities associated with its NGB nature. 	
From \Eq{eq:kinetic}, the kinetic term of the Higgs gets shifted after EWSB, which has the net effect of suppressing all of the Higgs interactions. 
On top of this, there are further corrections to the couplings to fermions, due to the non-standard Higgs dependence of their Yukawa couplings. 
Following the usual parametrization
\beq
\Lag_h =
\frac{h}{v} \left( a \left[ m_W^2 W_\mu^+ W^{- \mu} + m_Z^2 Z_\mu Z^\mu \right] - c_\psi^h m_{\psi} \bar{\psi} \psi \right) \ ,
\label{eq:lagh}
\eeq
one finds $a = \sqrt{1 - \xi}$, where $\xi = v^2/f^2$, and $c_\psi^h = (1-2\xi)/\sqrt{1-\xi}$, for both embeddings of $q_L$ and $t_R$ considered in this work, $\mathbf{6}_L + \mathbf{6}_R$ and $\mathbf{20}'_L + \mathbf{1}_R$. 
Notice however that these embeddings have only been identified for the top, and they need not be the same for the light quarks or the leptons. 
We have not included in \Eq{eq:lagh} the couplings of the Higgs to photons or gluons, since they are not significantly modified, beyond the rescaling of the top and $W$ loops induced by $a, c_\psi \neq 1$. 
In particular, light top partners do not give large contributions to such couplings in the models considered here \cite{Azatov:2011qy,Montull:2013mla}.
They do affect the Yukawa coupling of the top, by an amount of order $\lambda_{L,R}^2 v^2/m_\Psi^2$, which we will neglect to first approximation. 
Finally, the absence of mixing between the Higgs and the $\eta$ implies no further modifications of the Higgs couplings.\footnote{This is true at least for what regards the scalar potential generated by the third generation quarks and the $\SU(2)_L \times \UU(1)_Y$ gauge bosons, see section \ref{potential}.}

We can parametrize the linear couplings of $\eta$ in a similar fashion as those of the Higgs,
\bea
\Lag_\eta &=&
- i \frac{\eta}{v} \left( c_t^\eta m_{t} \bar{t} \gamma_5 t + c_b^\eta m_{b} \bar{b} \gamma_5 b + c_t^\eta m_{\tau} \bar{\tau} \gamma_5 \tau + c_c^\eta m_{c} \bar{c} \gamma_5 c \right) \nn \\
&& + \frac{\eta}{v} \left( c_g^\eta \frac{\alpha_s}{8 \pi} G_{\mu \nu} \tilde G^{\mu \nu} + c_\gamma^\eta \frac{\alpha}{8 \pi} A_{\mu \nu} \tilde A^{\mu \nu} \right) \nn \\
&& + \frac{\eta}{v} \left( c_{\gamma Z}^\eta \frac{\alpha}{8 \pi} A_{\mu \nu} \tilde Z^{\mu \nu} + c_Z^\eta \frac{\alpha}{8 \pi} Z_{\mu \nu} \tilde Z^{\mu \nu} + c_W^\eta \frac{\alpha}{8 \pi} W^+_{\mu \nu} \tilde W^{- \, \mu \nu} \right)
\, ,
\label{eq:lageta}
\eea
where we have now included couplings to the SM field strengths.
These are important for two reasons: first, given that $\eta$ is neutral under the electroweak interactions and it does not mix with the Higgs, the couplings to $W_\mu^+ W^{- \mu}$ and $Z_\mu Z^\mu$ vanish. 
Therefore the couplings to $F_{\mu \nu} \tilde F^{\mu \nu}$ are the leading ones to any of the SM gauge vectors.
And second, there can be direct contributions to this kind of couplings from UV physics, as explained in section \ref{bmchm}. 
Indeed we find from the anomalous term in \Eq{eq:anomaly}, 
\beq
c_g^\eta = n_g(2/3)\sqrt{\xi} = 0 \ , \quad c_\gamma^\eta = (n_W + n_B)(2/3)\sqrt{\xi} = 0 \ , \quad c_W^\eta = (n_W/\sin^2 \theta_W) (2/3)\sqrt{\xi} \ ,
\eeq
given $n_g = 0$ and $n_W = - n_B$. 
The rest of the couplings in \Eq{eq:lageta} are fixed by the relations $c_{\gamma Z}^\eta = (c_W^\eta - c_\gamma^\eta) \tan \theta_W$ and $c_Z^\eta = c_W^\eta - (c_W^\eta - c_\gamma^\eta) \tan^2 \theta_W$. 
Of course, given a non-vanishing coupling of $\eta$ to SM fermions, these can also contribute to the effective couplings to gluons and electroweak gauge bosons, much in the same way as they do for the Higgs (see for instance \cite{Gunion:1989we}). 
In order to fix the coefficients $c_\psi^{\eta}$ in \Eq{eq:lageta}, we must specify the embeddings of $\psi = t, b, \tau, c$. 
Let us assume that for the bottom, tau, and charm, these are the same than for the top, that is either $\mathbf{6}_L + \mathbf{6}_R$ or $\mathbf{20}'_L + \mathbf{1}_R$ (we rename $\hat \gamma \equiv \gamma$ for notational simplicity in this section). 
In both cases we find 
\beq
c_i^\eta = \gamma_i \sqrt{\xi} \ , \ i = t, b, \tau, c \ ,
\eeq
where we recall $\xi = v^2/f^2$. 
For what regards the values of the different $\gamma$'s, let us recall that in the limit $\gamma_i \to 1$ the $\UU(1)_\eta$ symmetry is unbroken and the singlet does not get a potential from loops of the fermion $i$. 
Let us also notice that the limit $\gamma_i \to 0$ is associated to a $P_\eta$ parity symmetry under which the fermion $i$ is even and $\eta$ is odd. 
This just means that there are selection rules which we can use to naturally take either $\gamma_i = \order{1}$ or $\gamma_i \ll 1$. 

\begin{figure}[!t]
\begin{center}
\includegraphics[width=3.1in]{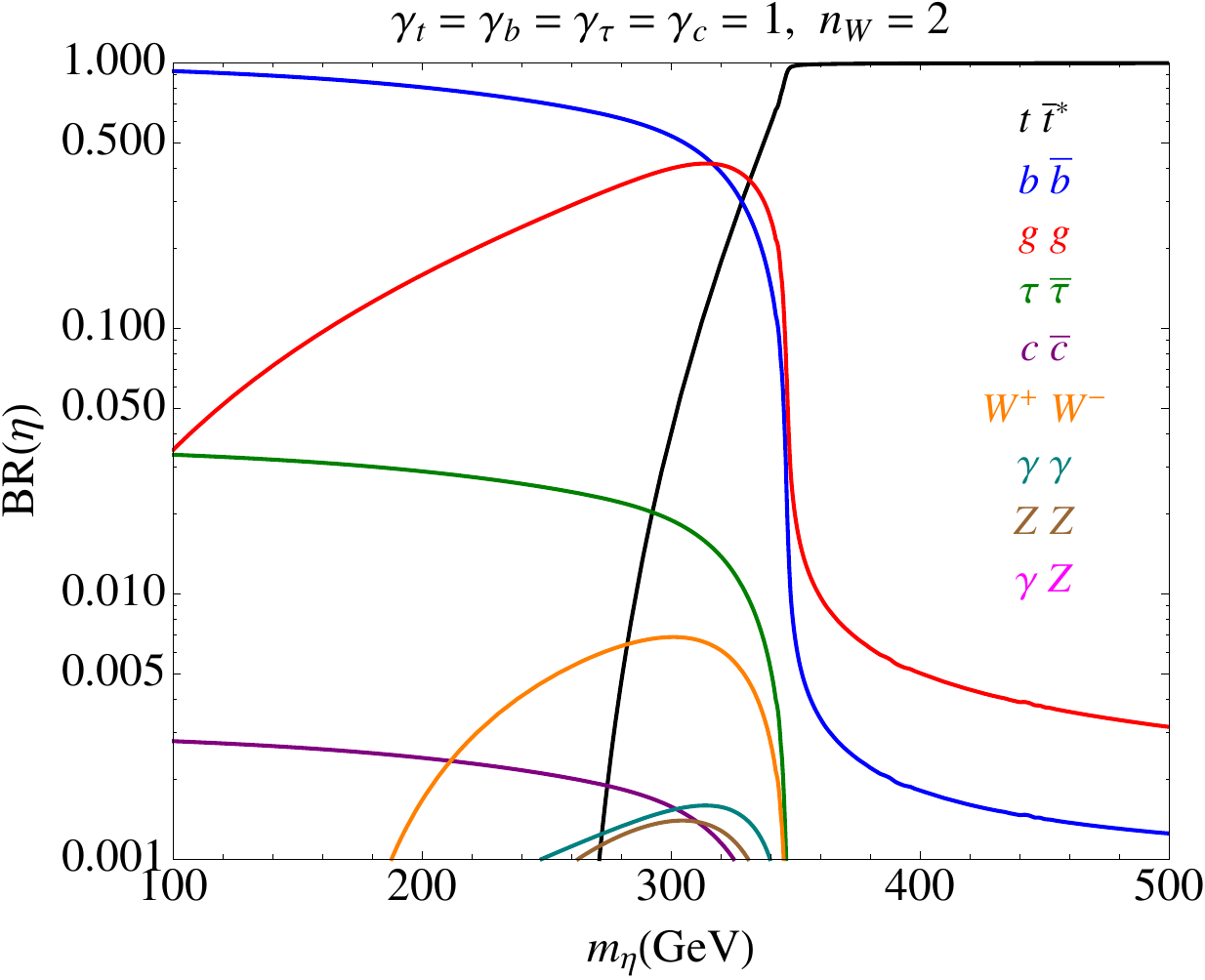}
\hspace{7pt}
\includegraphics[width=3.1in]{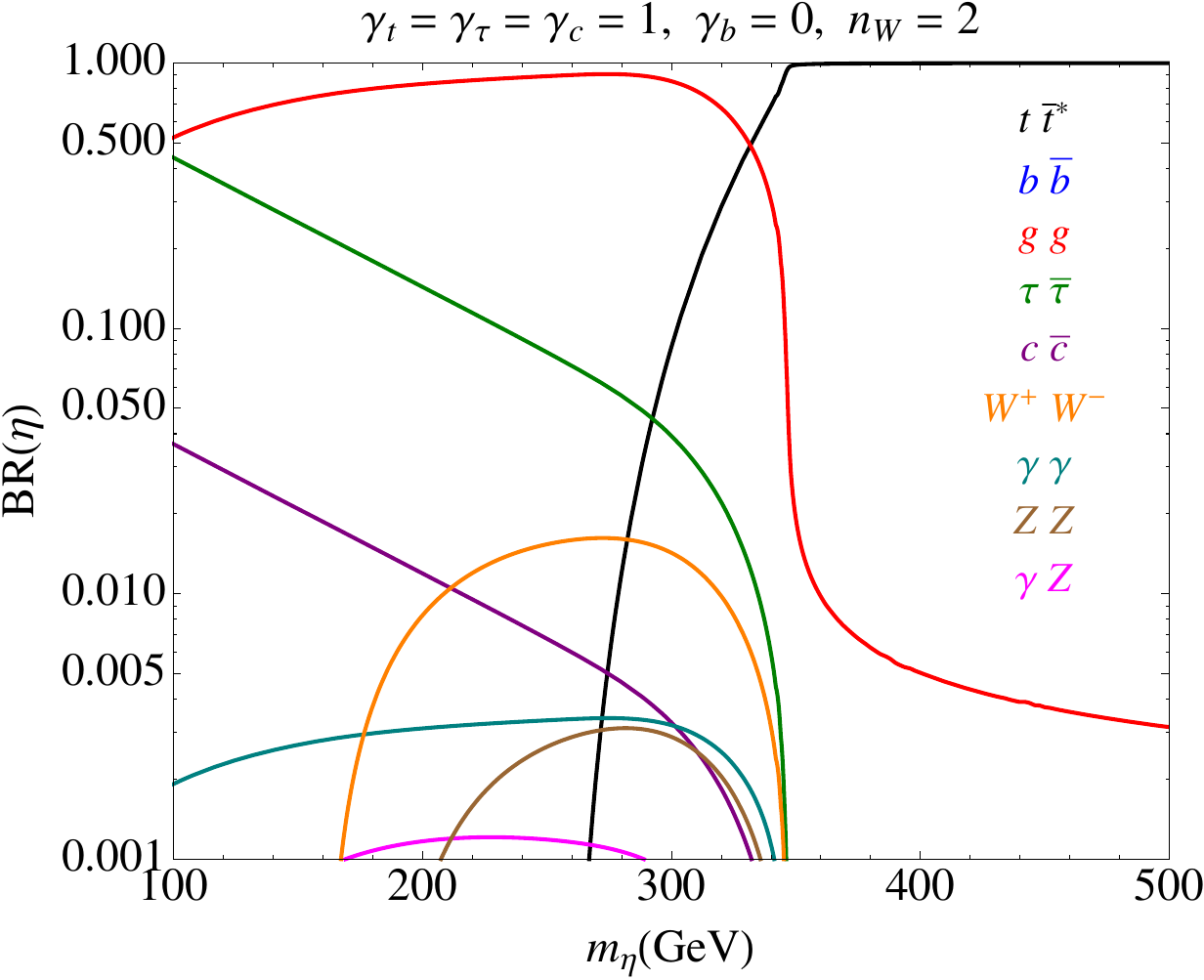}
\caption{Branching ratios of $\eta$ to $t \bar t^*$ (black), $b \bar b$ (blue), $gg$ (red), $\tau \bar \tau$ (green), $c \bar c$ (purple), $W^+ W^-$ (orange), $\gamma \gamma$ (turquoise), $ZZ$ (brown), and $\gamma Z$ (magenta). 
The left panel correspond to universal couplings to SM fermions $\gamma_t = \gamma_b = \gamma_\tau = \gamma_c = 1$, while in the right pane we suppressed the coupling to bottoms, $\gamma_b = 0$. 
The anomaly coefficient has been fixed in both cases to $n_W = 2$.}
\label{fig:breta}
\end{center}
\end{figure}

All the $\eta$ couplings carry a $\sqrt{\xi}$ suppressing factor. 
Consequently, for $\gamma_i = 1$ the single production cross-sections of the singlet are the same as those of a SM Higgs of mass $m_\eta$, times a $\xi$ factor. 
This of course excludes all processes involving the electroweak gauge bosons, \ie vector boson fusion and Higgs-strahlung. 
The suppression due to $\xi \lesssim 0.2$ significantly reduces the production rate of $\eta$'s. 
For what regards the branching ratios, the factor of $\xi$ drops out (they do not depend on $f$). 
Therefore, for $\gamma_i = 1$ and keeping in mind that $\eta$ does not couple linearly to the longitudinal $W^\pm$ and $Z$, the $\BR$'s of the singlet should follow the same pattern as those of the SM Higgs. 
This is explicitly shown in the left panel of Figure~\ref{fig:breta}. 
There we neglected the contributions from loops of SM fermions to the couplings with $\gamma Z$, $ZZ$, and $W^+ W^-$. 
These are generically subleading, and moreover they are the only ones that receive a contribution directly from the UV anomalies, for which we took $n_W = 2$. 
Since the couplings to tops is much larger than the rest, we also included off-shell top effects in the decay to $t \bar t$ (see for instance \cite{Djouadi:1995gv}). 
From the left panel of Figure~\ref{fig:breta} we can then conclude that for $\order{1}$ couplings to SM fermions, $\eta$ mostly decays to bottom pairs below the $t \bar t$ threshold, while above it decays to top pairs. 
The situation changes significantly if $\gamma_b \ll 1$ while the other Yukawa couplings are order one. 
Below the $t \bar t$ threshold $\eta$ mostly decays to gluons in that case. 
This is shown in the right panel of Figure~\ref{fig:breta}. 
There are several other situations that we could consider, exposing the variability of the phenomenology of the pseudo-scalar singlet.  
When $\gamma_t > 1$, for which the decay $\Psis \to \eta t$ is enhanced, the decay of $\eta$ to gluons is enhanced because of the larger contribution from the top loop, and dominates over $b \bar b$ at low masses (low in the sense of below the $t \bar t$ threshold). 
If $\gamma_b \ll 1$ but $\gamma_\tau \gg 1$, the $\BR(\eta \to \tau \bar \tau)$ is enhanced and dominates over that to gluons, thus the singlet becomes a $\tau \bar \tau$ resonance at low masses. 
And if both $\gamma_b, \gamma_\tau \ll 1$ while $\gamma_c \gg 1$, then $\eta$ becomes a $c \bar c$ resonance, or in other words it decays mostly to jets. 
Finally, when $\gamma_t = 0$ and the rest of Yukawa couplings are order one, the $\BR(\eta \to b \bar b)$ dominates over the whole mass range. 

Notice that in the discussion above we have assumed that the couplings of $\eta$ to fermions respected $\cp$, that is $\gamma_i \in \mathbb{R}$. 
If that was not the case, a tadpole term would be induced for $\eta$, which nevertheless would be proportional to $y_i^2 \Im[\gamma_i] \Re[\gamma_i]$, thus small and under control. 
Furthermore, let us recall that the predictions for $m_\eta$ in section \ref{potential} were close to the $t \bar t$ threshold, implying that both possibilities $m_\eta \lessgtr 2 m_t$ should be equally considered. 
However, we do not contemplate here the case in which the singlet is light enough for the Higgs to decay to $\eta \eta$ (we refer to \cite{Falkowski:2013dza} where this possibility is partly discussed in the light of the Higgs discovery). 
Finally, let us notice that given the prospect of top partners decaying significantly to $\eta t$, their production could become an important source of $\eta$'s. 
Another extra production mechanism for the pseudo-scalar could proceed via a large coupling to bottom quarks, $\gamma_b \gg 1$, boosting production through or in association with bottom quarks.


\section{Outlook} \label{conclusions}

Top partners are expected to be the first sign of new physics associated to the naturalness problem of the electroweak scale, both in composite Higgs models and in supersymmetric extensions of the SM. 
In this work we have investigated the role of the top partners in the Next to Minimal Composite Higgs Model. 
These fermionic resonances, related to the top quark, control the size of the Higgs potential by effectively cutting off the radiative contributions associated to the top Yukawa coupling. 
We have explicitly shown that in the NMCHM, keeping fine tuning to the minimum and reproducing the Higgs mass requires the top partners to be light and weakly coupled, aspect shared with most models. 

One of the characteristic features of the NMCHM is the presence in the spectrum of a light pseudo-scalar $\eta$, singlet under the SM gauge symmetries.
This arises as a Nambu-Goldstone boson along with the Higgs from the spontaneous breaking of a global $\SO(6)$ symmetry down to $\SO(5)$. 
Interestingly, the decay patterns of the top partners can be significantly affected by this extra state. 
We have identified under which conditions the decays of $\Psis$, a top partner singlet of $\SO(5)$, are dominated by the $\eta t$ channel. 
We have also shown that certain exotic top partners in the 5-plet of $\SO(5)$, which arise from the extended symmetry structure of the NMCHM, decay to $\eta t$ only. 
Motivated by the preference, in the simple models studied here, for a singlet top partner lighter than the 5-plet, we have discussed as well the feasibility of the decays $\Psif \to \Psis \Pi$, $\Pi = W^\pm, Z, h, \eta$.
In addition, we have explicitly verified with several examples the viability of all such non-standard decays with respect to the generation of the Higgs potential. 
It is worth noting that while the NMCHM is the simplest extension of the minimal composite Higgs model with custodial protection \cite{Agashe:2004rs}, there is a plethora of other possibilities for the quantum numbers of non-minimal NGBs, which could play a similar role in top partner decays \cite{Bellazzini:2014yua}. 

One question we have left unanswered in this work is how much the experimental bounds on the top partners change given $\BR(\Psi \to \eta t) \neq 0$. 
If we simply take the extra decay channel as a reduction of the standard branching ratios ($h t$, $W^\pm b$, $Z t$), then we roughly estimate that the bounds could go down as much as $\sim 100 \GeV$ for the singlet $\Psis$, while they would be absent for those top partners that decay exclusively to $\eta t$. 
However, the experimental searches could be recasted or adjusted to look for the different pattern of final state particles from the production and decay of these top partners.
We expect that the corresponding analyses could reach comparable sensitivities as the current ones (see \cite{Leskow:2014kga,Anandakrishnan:2015yfa} where this subject is addressed). 
Nevertheless, the search for non-minimal top partner decays could also provide compelling information about the underlying symmetry structure of the electroweak scale. 
From another point of view, analyses incorporating inclusive decays such as $\Psi \to t + X$ would certainly contribute to cover most of the ground regarding detection of top partners at colliders, much in the same way as the study of non-standard Higgs decays has been carried out \cite{Gunion:1989we}. 

The mass of the pseudo-scalar singlet is predicted to be a factor $\sim f/v$ larger than that of the Higgs. 
Above the $t \bar t$ threshold $\eta$ decays almost with branching ratio one to top pairs, while for lower masses its decays are more model dependent. 
When the coupling to bottoms is unsuppressed, $\eta \to b \bar b$ dominates. 
Instead, if the singlet does not couple to bottoms, detection at colliders becomes challenging, since it mostly decays to pairs of jets. 
Still, the coupling of $\eta$ to taus could be enhanced, in which case $\eta \to \tau \bar \tau$ would become the dominant decay channel, and likewise for $\eta \to c \bar c$. 
The singlet is mostly produced through gluon fusion, although with a cross section suppressed by $v^2/f^2$. 
It is important to remark that $\eta$ generically presents a phenomenology substantially different than that of an elementary (pseudo-)scalar singlet. 

In conclusion, the NMCHM is a simple non-minimal composite Higgs model which presents a top partner phenomenology that is non-standard, while retaining experimental consistency with little tuning.


\section*{Acknowledgements}

I would like to thank Clara Peset, Riccardo Torre and Andrea Wulzer for helpful discussions and valuable comments on the manuscript. 
This work was supported in part by the MIUR-FIRB Grant RBFR12H1MW.


\appendix

\section{Explicit representations} \label{ccwz}

In the vectorial $\mathbf{6}$ representation of $\SO(6)$, we have chosen the generators as
\begin{align}
& (T^{\alpha}_{L})_{IJ} = -\frac{i}{2} \left[ \frac{1}{2} \epsilon^{abc} \left(  \delta^b_I \delta^c_J - \delta^b_J \delta^c_I \right) + \left(  \delta^a_I \delta^4_J - \delta^a_J \delta^4_I \right) \right] \ , \ \alpha = 1, 2, 3 \ , \nonumber \\
& (T^{\alpha}_{R})_{IJ} = -\frac{i}{2} \left[ \frac{1}{2} \epsilon^{abc} \left(  \delta^b_I \delta^c_J - \delta^b_J \delta^c_I \right) - \left(  \delta^a_I \delta^4_J - \delta^a_J \delta^4_I \right) \right] \ , \ \alpha = 1, 2, 3 \ , \nonumber \\
& (T^{\beta}_{\mathbf{1}})_{IJ} = - \frac{i}{\sqrt{2}} \left(  \delta^{i}_I \delta^5_J - \delta^{i}_J \delta^5_I \right) \ , \ \beta = 1, \ldots, 4 \ , \nonumber \\
& (T^{\beta}_{\mathbf{2}})_{IJ} = - \frac{i}{\sqrt{2}} \left(  \delta^{i}_I \delta^6_J - \delta^{i}_J \delta^6_I \right) \ , \ \beta = 1, \ldots, 4 \ , \nonumber \\
& (T_{\eta})_{IJ} = - \frac{i}{\sqrt{2}} \left(  \delta^5_I \delta^6_J - \delta^5_J \delta^6_I \right) \ , 
\label{eq:gen}
\end{align}
where $I, J = 1, \ldots, 6$. 
The $\SO(5)$ unbroken generators are identified with $T^a = \{ T^{\alpha}_{L}, T^{\alpha}_{R}, T^{\beta}_{\mathbf{1}} \}$, while the $\SO(6)/\SO(5)$ broken generators are $T^i = \{ T^{\beta}_{\mathbf{2}}, T_{\eta} \}$. 
The generators $T^{\alpha}_{R,L}$ span the custodial $\SO(4) \cong \SU(2)_L \times \SU(2)_R$ subgroup of $\SO(5)$, while $T_{\eta}$ is the extra Cartan generator, corresponding to the $\UU(1)_\eta$ abelian symmetry. 
The SM electroweak symmetry group is identified with the generators of $\SO(6)$ as $T^a_{L} = T^{\alpha}_{L}$ and $Y = T^{3}_{R}$. 

From the Goldstone matrix $U(\Pi(x))$ we can construct the $d$ and $e$ symbols \cite{ccwzref},
\beq
- i U^\dagger D_\mu U = d_\mu^i T^i + e_\mu^a T^a \equiv d_\mu + e_\mu \ ,
\label{eq:form}
\eeq
which transform as
\bea
d_\mu \!\!\! & \to & \!\!\! \hat h(\Pi,g) d_\mu \hat h^\dagger(\Pi,g) \ , \\
e_\mu \!\!\! & \to & \!\!\! \hat h(\Pi,g) e_\mu \hat h^\dagger(\Pi,g) - i \hat h(\Pi,g) \partial_\mu \hat h^\dagger(\Pi,g) \ ,
\label{eq:detrans}
\eea
with $g$ a global $\SO(6)$ transformation and $\hat h$ a local (dependent on $\Pi(x)$) $\SO(5)$ transformation. 
Given that the SM subgroup of $\SO(6)$ is gauged, we must also consider local $g$ transformations. 
These are incorporated through $D_\mu = \partial_\mu - i A_\mu$ in \Eq{eq:form}, where $A_\mu = A_\mu^a T^a = g W_\mu^a T^a_{L} + g' B_\mu Y$.
At lowest order in the NGBs, the $d$ and $e$ symbols then read
\beq
d_\mu^i = \frac{\sqrt{2}}{f} D_\mu \Pi^i + \cdots \ , \quad e_\mu^a = - A_\mu^a + \cdots \ .
\label{eq:defirst}
\eeq

The $\SO(5)$ multiplets of top partners introduced in sections \ref{potential} and  \ref{pheno} transform as
\beq
\Psis \to \Psis \ , \quad \Psif \to \hat h(\Pi, g) \Psif \ , \Psi^{\mathbf{14}} \to \hat h(\Pi, g) \Psi^{\mathbf{14}} \hat h^T(\Pi, g) \ .
\eeq

Finally, an explicit representation for the top partners $\Psi$ in the $\mathbf{20'}$ of $\SO(6)$ is given by
\beq
\Psi = 
\begin{pmatrix}
   \Psi^{\mathbf{14}} - 1_{5 \times 5} \, \Psis/\sqrt{30}  & \Psif/\sqrt{2}  \\
    (\Psif)^T/\sqrt{2}   & \sqrt{5/6} \, \Psis \\  
\end{pmatrix} \ ,
\label{eq:psi20}
\eeq
where $\Psif$ is written as in \Eq{eq:psi6}, while $\Psi^{\mathbf{14}}$ is a symmetric trasceless tensor, which further decomposes as a $\mathbf{1} + \mathbf{4} + \mathbf{9}$ of $\SO(4)$,
\beq
\Psi^{\mathbf{14}} = 
\begin{pmatrix}
   \Psi^{\mathbf{9}} - 1_{4 \times 4} \, \tilde T'/ 2 \sqrt{5}  & \Psi^{\mathbf{4}}/\sqrt{2}  \\
    (\Psi^{\mathbf{4}})^T/\sqrt{2}   & 2 \, \tilde T' /\sqrt{5} \\  
\end{pmatrix} \ ,
\label{eq:psi14}
\eeq
with
\beq
\Psi^{\mathbf{4}} = \frac{1}{\sqrt{2}}
\begin{pmatrix}
      - (\tilde B - \tilde X_{5/3})    \\
      i (\tilde B + \tilde X_{5/3})    \\
      - (\tilde X_{2/3} + \tilde T)    \\
      i (\tilde X_{2/3} - \tilde T)    \\
\end{pmatrix} \ ,
\eeq
and
\beq
\Psi^{\mathbf{9}} = 
{\tiny
\frac{1}{2}
\begin{pmatrix}
      X_{{8 \over 3}}^{+} + i X_{- {4 \over 3}}^{-} - X_{{2 \over 3}}^{0} \hspace{-3pt} & X_{- {4 \over 3}}^{-} + i X_{{8 \over 3}}^{+} & \frac{1}{\sqrt{2}} \left( X_{{5 \over 3}}^{0} + i X_{{5 \over 3}}^{+} + X_{-{1 \over 3}}^{-} + i X_{-{1 \over 3}}^{0} \right) \hspace{-3pt} & \frac{1}{\sqrt{2}} \left( i X_{{5 \over 3}}^{0} + X_{{5 \over 3}}^{+} + i X_{-{1 \over 3}}^{-} + X_{-{1 \over 3}}^{0} \right) \\
     & - X_{{8 \over 3}}^{+} - i X_{- {4 \over 3}}^{-} - X_{{2 \over 3}}^{0} \hspace{-3pt} & \frac{1}{\sqrt{2}} \left( i X_{{5 \over 3}}^{0} - X_{{5 \over 3}}^{+} - i X_{-{1 \over 3}}^{-} + X_{-{1 \over 3}}^{0} \right) & \frac{1}{\sqrt{2}} \left( i X_{{5 \over 3}}^{+} - X_{{5 \over 3}}^{0} - i X_{-{1 \over 3}}^{0} + X_{-{1 \over 3}}^{-} \right) \\
       &  & X_{{2 \over 3}}^{-} - X_{{2 \over 3}}^{+} + X_{{2 \over 3}}^{0} & i \left( X_{{2 \over 3}}^{-} + X_{{2 \over 3}}^{+} \right) \\
       &  &  & X_{{2 \over 3}}^{+} - X_{{2 \over 3}}^{-} + X_{{2 \over 3}}^{0} \\     
\end{pmatrix} \ .
}
\eeq
The notation for the components of $\Psi^{\mathbf{9}}$ has been chosen such that the states with the same upper index belong to the same $\SU(2)_L$ multiplet.


\bibliographystyle{unsrt}

\end{document}